\documentclass[twocolumn,secnumarabic,amssymb, nobibnotes, aps, prd, unsortedadress]{revtex4-1}
\usepackage{siunitx}
\usepackage{graphicx}
\usepackage{amsmath}
\usepackage{xcolor}
\usepackage{hyperref}
\usepackage{comment}

\begin{document}

\title{The signatures of secondary leptons in radio-neutrino detectors in ice}%

\author{D.~Garc\'{\i}a-Fern\'{a}ndez}
\email[]{daniel.garcia@desy.de}
\affiliation{DESY, Platanenallee 6, 15738, Zeuthen, Germany\\
ECAP, Friedrich-Alexander Universität Erlangen-Nürnberg, Erwin-Rommel-Straße 1, 91058 Erlangen, Germany}
\author{C.~Glaser}
\email[]{christian.glaser@physics.uu.se}
\affiliation{University of California, Irvine, CA 92697, Irvine, USA \\ Uppsala University, Box 516, S-75120 Uppsala, Sweden}
\author{A.~Nelles}%
\email[]{anna.nelles@desy.de}
\affiliation{DESY, Platanenallee 6, 15738, Zeuthen, Germany\\
ECAP, Friedrich-Alexander Universität Erlangen-Nürnberg, Erwin-Rommel-Straße 1, 91058 Erlangen, Germany}

\begin{abstract}
The detection of the radio emission following a neutrino interaction in ice is a promising technique to obtain significant sensitivities to neutrinos with energies above PeV. The detectable radio emission stems from particle showers in the ice. So far, detector simulations have considered only the radio emission from the primary interaction of the neutrino. For this study, existing simulation tools have been extended to cover secondary interactions from muons and taus. We find that secondary interactions of both leptons add up to 25\% to the effective volume of neutrino detectors. Also, muon and tau neutrinos can create several detectable showers, with the result that double signatures do not constitute an exclusive signature for tau neutrinos. We also find that the background of atmospheric muons from cosmic rays is non-negligible for in-ice arrays and that an air shower veto should be considered helpful for radio detectors. 
\end{abstract}

\maketitle

\section{Introduction}
The flux of astrophysical neutrinos at PeV energies is by now firmly established \cite{Aartsen:2015rwa}. With these energies in reach, attention is turning to energies beyond tens of PeV up to EeV energies. At these energies, flux predictions for neutrinos from sources and neutrinos generated from propagating ultra-high energy cosmic rays vary greatly \cite{Ackermann:2019ows}. It is however likely that detectors much bigger than current ones are needed to obtain the necessary sensitivities. Even expanding the optical detector IceCube by a factor of 10 may not be large enough. Thus, the community is considering building detectors that target the radio emission following a neutrino interaction, thereby exploiting the kilometer-scale attenuation length of radio emission in ice. 

Early stage pathfinder arrays, such as ARA \cite{Allison:2018ynt} and ARIANNA \cite{Anker:2019mnx} have shown that radio detection is feasible and faces no general show-stoppers. The upcoming projects, such as RNO-G, scheduled to start deployment in summer 2021, will now have to show that the technique can be scaled up to finally be included in the future extension of IceCube, IceCube-Gen2 \cite{Aartsen:2019swn}. 

With experimental efforts reaching maturity, the software and simulation tools have to follow along. So far, simulation tools focused on calculating effective volumes for different types of neutrino detectors, thereby including only the primary neutrino interaction. Also theoretical works have been published that include general considerations of the tau decay as well. Recently, a new framework, NuRadioMC, was set up to be able to flexibly simulate neutrino detectors and optimize their layouts in preparation for larger scale deployments \cite{NuRadioMC}. Simulations are, however, not only needed to establish the sensitivities of the detector, but also to estimate potential backgrounds and secondary detection channels. We have thus extended NuRadioMC to study contributions of interactions from secondary leptons, which is presented in this article. This is the first time the calculation of radio emission from neutrino interaction to a detector in ice contains in a systematic way all the possible showers produced by secondary leptons.

This paper is structured as follows: We first discuss the relevant physical processes in Sect.~\ref{sec:processes}, such as the nature of the radio emission and the particle physics involved in interactions of secondary leptons. In Sect.~\ref{sec:simulations}, we describe the changes to NuRadioMC and the detector layouts that have been used for this study. This is followed by a study of the background induced by atmospheric muons in Sect.~\ref{sec:atmospheric}. Finally, we show the contribution of interactions of secondary leptons to the neutrino effective volumes of radio detectors in Sect.~\ref{sec:effective_volumes}. We also show event topologies and discuss multiple signatures resulting from the same neutrino interaction. All sections discuss practical implications for future radio neutrino detectors.

\section{Physical processes in the interaction of neutrinos}
\label{sec:processes}
In this section we will discuss both the nature of the radio emission following a neutrino interaction and the particle physics relevant to create showers from secondary leptons that lead to radio emission.

\subsection{Radio emission of particle showers}

When a high-energy particle interacts within a medium, a particle shower is created. The charged particles in this shower create a radiation electric field as they propagate. In principle, particle interactions in vacuum are approximately charge-symmetric and the shower should have almost zero charge, thus there would be no coherent electric field emission from the shower. However, when the particle shower develops in a medium and it accumulates electrons from the medium, a process that is called the Askaryan effect \cite{Askaryan:1962hbi}, the total charge of the shower becomes negative. Besides, the geomagnetic field deviates opposite particles in opposite directions, creating an overall electric current \cite{KahnLerche}.

Both the emission due to the net negative charge (due to the Askaryan effect) and the geomagnetic effect create a coherent emission at low frequencies (down to tens of MHz). Since the shower travels at a speed faster than the speed of light in the medium it develops, Cherenkov-like relativistic effects are present. This implies that if the shower is observed near the Cherenkov angle, the emitted electric field from a large portion of the shower will be observed at the same time, which will increase coherence up to GHz frequencies \cite{ZHAireS-Reflex}. 
In dense media such as ice, which is the medium we will consider throughout this paper, the Askaryan effect is dominant. Due to the high density, the particle shower develops in less space (at most tens of meters compared to several kilometers in air) and the deviation induced by the geomagnetic field is small. For a $\SI{1}{GeV}$ electron traveling perpendicular to a $\SI{25}{\micro\tesla}$ geomagnetic field, the Larmor radius is $\SI{\sim 6.7e6}{m}$, much larger than the shower dimensions in ice, making the geomagnetic effect negligible for dense media showers.

In general, the problem of radio emission from particle showers is a complex one that depends on the medium, shower size and particle distribution, observation frequency, and observer position. It must be calculated microscopically using Monte Carlo codes \cite{ZHAireS,CoREAS,selfas3}, if high accuracy is needed. However, macroscopic models can reproduce the overall characteristics of the emission and help us gain insight \cite{MGMR,EVA,NuRadioMC}. Alternatively, numerical methods like the finite differences in time domain (FDTD) method can be used in special cases to calculate the electric field from particle showers \cite{chen}.

The most accurate way of calculating the electric field for a shower developing in a dense homogeneous medium is simulating the particle tracks conforming the particle shower and calculating the electric field as the sum of all the fields produced by the particle tracks. This is the approach of the ZHS code \cite{zhs}. The result from this Monte Carlo code can be approximated by a set of parameterizations for hadronic and electromagnetic showers \cite{Alvarez2000,Alvarez2009} where the latter model accounts for the for the  Landau-Pomeranchuck-Migdal (LPM) effect \cite{Landau:1953um,Migdal:1956tc} on a stochastic basis via an effective stretching of the longitudinal shower profile. This model is used throughout the paper as it provides an accurate description of the electric-field amplitudes at negligible computing time \cite{NuRadioMC}. We note that also a semi-analytical time-domain model exists \cite{ARZ,Alvarez-Muniz:2020ary}, which improves the prediction of the pulse shape, modeling of LPM showers and agrees within 3\% with a microscopic Monte Carlo simulation but comes at the expense of increased computing time. Both models have been implemented in NuRadioMC and for effective volume calculations the differences are small, with larger differences being observed only for individual events. Thus, for studies of general characteristics, with an eye on computing resources, we default to using the signal parameterization of \cite{Alvarez2009}.   

Big natural ice volumes near the surface are not homogeneous \cite{southpole_2005,greenland_att}. To simulate real detectors, wave propagation in a non-homogeneous medium must be taken into account and the ZHS Monte Carlo or its parameterizations do not suffice \footnote{Even for antennas deep in the ice, surface effects play a role when signals are reflected or refracted}. A way of tackling this problem is to disentangle signal generation
from signal propagation and use a ray-tracing algorithm for calculating the trajectory of the electromagnetic wave.
We will use NuRadioMC for our calculations, which incorporates a ray-tracing module \cite{NuRadioMC}.
The model for the refractive index is an exponential function that depends on the vertical coordinate in the ice, which is a good description of the available ice-density measurements \cite{southpole_2005, alley_1988, hawley_2008} and it offers
the advantage that the ray tracing solutions can be analytically calculated \cite{NuRadioMC}. The frequency-dependent 
attenuation of the radio waves \cite{mb1, greenland_att} is also taken into account.
Focusing corrections due to the bending (focusing) of signal trajectories in the firn are also included in NuRadioMC, although its influence in the overall effective volume seems to only be important for low energies and inducing a correction of $\sim 10\%$ at high energies.

The typical electric field created by a particle shower in ice is a bipolar pulse that lasts for a few nanoseconds (see \cite{NuRadioMC} for an example). No instrument, however, measures the true pulse, due to limited bandwidth and sampling. In simulations, this electric field is thus convolved with the antenna response and the response of the electronics (amplifiers, filters, etc.), which results in the waveform as recorded in the detector. If the waveform is recorded in many antennas and the signal-to-noise ratio is large enough, the properties of the electric field emitted by the particle shower can be reconstructed. The electric field then allows to extrapolate properties of the primary particle, as for example shown in \cite{glaser_icrc} or in \cite{Welling_2019} for cosmic rays.
 
There are simple relations that can be used to qualitatively understand the radio emission. The pulse amplitude scales linearly with the energy of the shower (ignoring the LPM effect), meaning that at the same observing distance the pulses can cover many decades of size. In practice, however, neutrino interactions are scarce so the observable volume scales with the energy of the shower, thus, the most probable detection of signals is still near the threshold. The threshold is determined by the energy at which the pulse amplitude is above the noise level of the system, so for close showers typically above 1 PeV or higher. 

\subsection{Relevant particle physics}

When a neutrino undergoes a charged current (CC) interaction, a lepton and a hadronic cascade are produced. In the case of the electron neutrino, an electron is produced along with the hadronic cascade, which then almost immediately creates an electromagnetic shower at the same place. The radio emission from two showers interfere, and at low energies, for which the LPM effect is negligible, the shower maxima of both showers are close and the interference is mostly constructive. If the LPM effect is relevant, the electromagnetic shower can become very elongated and its maximum can be far away from the hadronic shower maximum, so the showers can interfere destructively or they can be seen as two (or more) independent pulses \cite{Alvarez-Muniz:2020ary,alvarez_lpm,ARZ}. Due to the probability distribution of the inelasticity (i.e. how much energy is transferred to the electron versus the hadronic shower), the typical interaction will have the majority of the neutrino energy transferred to the electron, for which interference effects are small. 

In case a muon neutrino interacts via CC, a muon is produced, while for tau neutrinos, a tau is produced.
The muon and the tau continue their propagation until they decay. Muons lose most of their energy through bremsstrahlung, pair production, and nuclear interactions, and they typically have low energies upon decay. This means that muons are susceptible to produce subsequent showers if they radiate, for instance, a bremsstrahlung photon or hadrons above a certain energy. If the shower energy is above the experiment-dependent radio detection threshold, it can be detected.
Taus radiate electron-positron pairs mainly, which creates electromagnetic cascades, but they tend to lose larger amounts of energy via photonuclear interaction, which creates hadronic cascades. Taus can also decay while they have large energies, into hadronic and leptonic channels. These decays produce showers and depending on the channel, muons can be produced, which in turn may produce more showers \cite{Dutta:2000hh}.

The initial tau neutrino-induced shower, together with a shower produced upon tau decay, constitutes the long sought-after \textit{double-cascade} neutrino signature reported by IceCube \cite{Stachurska:2019wfb}. While it is expected to be a detectable signature also for the radio technique \cite{double_bang}, no existing in-ice radio emission codes accounts for tau decay in a rigorous way. And although the showers induced by leptons in dense media have been studied for long (see, for instance, \cite{lipari_muons}), the study of the radio emission created by lepton-induced showers associated to a neutrino event due to radiative losses has never been carried out. 

When detecting in-ice cascades, a phased array configuration like the one in \cite{Allison:2018ynt} can have a detection threshold as low as \SI{\sim 1}{PeV} in shower energy, which is considered most ambitious. Therefore, throughout this work, we will consider \SI{1}{PeV} as the minimum energy a shower must possess to be detectable using radio. 
The threshold of \SI{1}{PeV} for particle showers is not an arbitrary choice. Radio detection of showers has a 'natural' threshold, as the shower must emit enough radiation to induce a pulse that can be detected above the thermal noise caused by
the temperature of the ambient medium, the electronics chain, and the Galaxy. Depending on the characteristics of the detector, the threshold
can go down to a few PeV but it is considered extremely challenging to detect showers below \SI{1}{PeV}. The pulse amplitude scales linearly with energy
and with the inverse of the distance, which means that a \SI{2}{PeV} shower creates a pulse twice as large as a \SI{1}{PeV}
shower, at the same distance. Choosing an initial \SI{1}{PeV} threshold together with a sensible trigger scheme ensures that we are not
ignoring any shower that could trigger our system.

\subsection{Shower-inducing secondaries from taus and muons}

\begin{figure}
    \centering
    \includegraphics[width=0.48\textwidth]{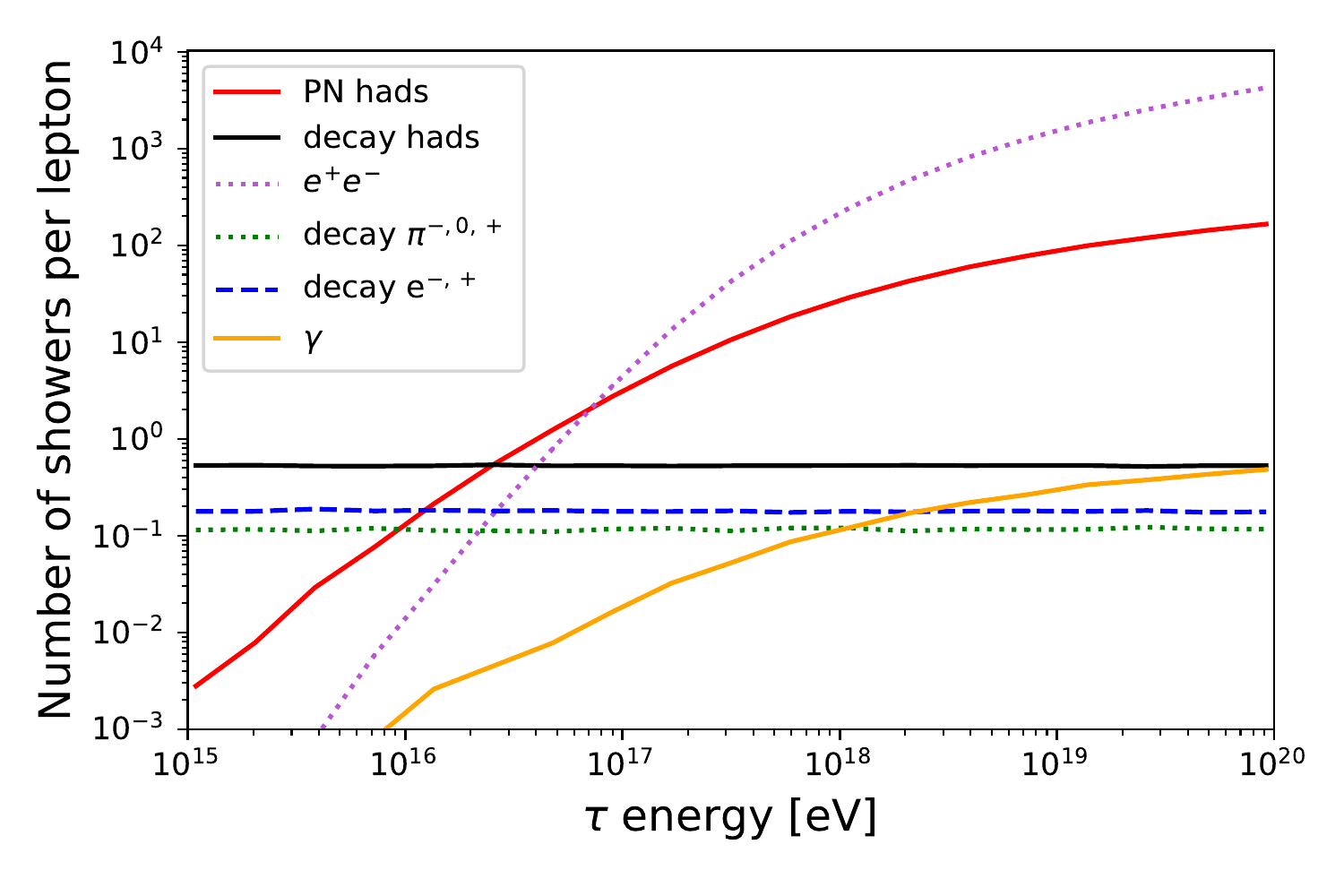}
    \includegraphics[width=0.48\textwidth]{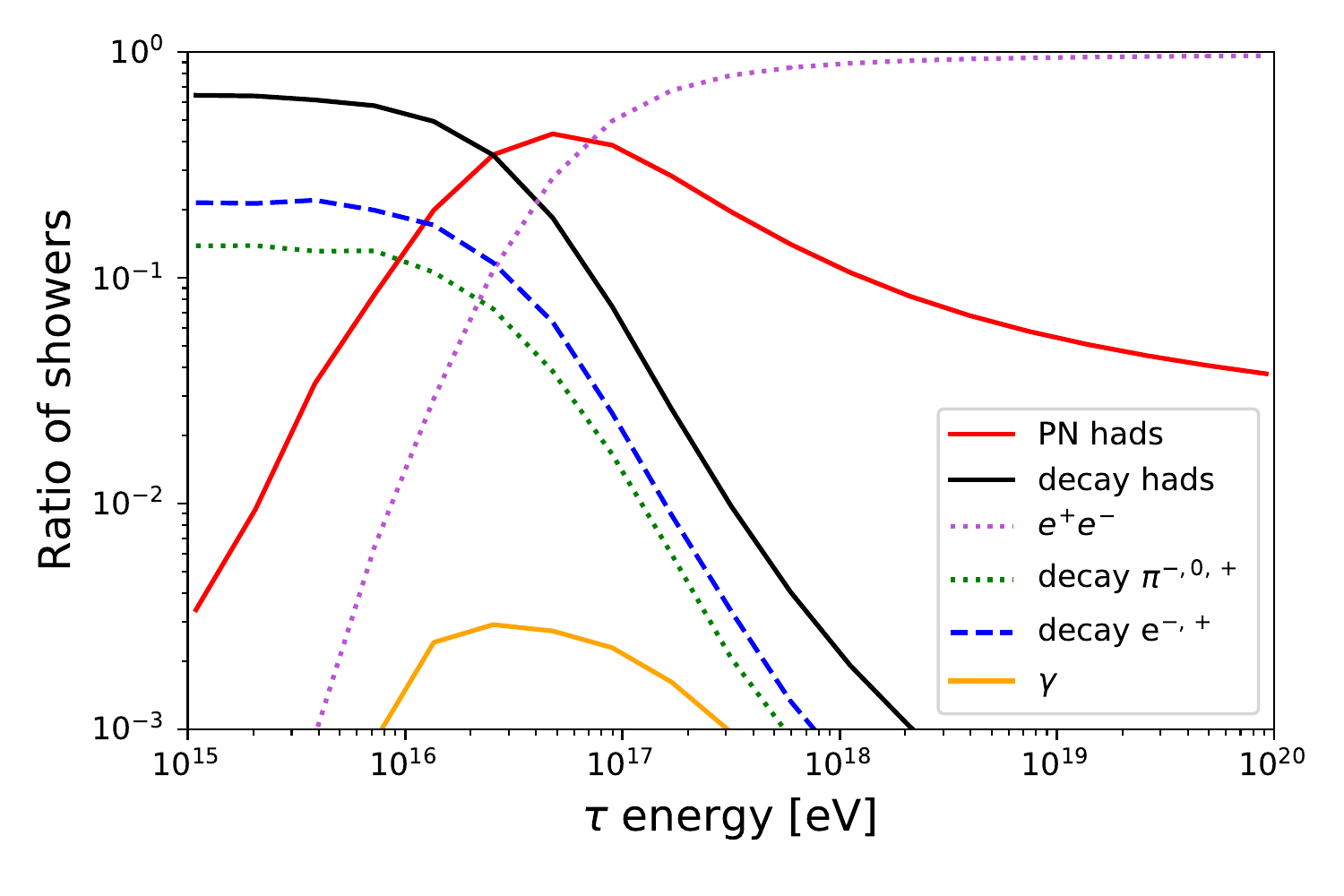}
    \caption{Top: Average number of ($>\SI{1}{PeV}$) showers produced by a tau as a function of the initial energy, classified by shower-initiating secondary particle type. Bottom: Ratio of average number of showers per primary type over total number of particle showers for a tau. The shower primaries in the legend are noted as follows. Decay hads: hadron bundle created upon decay. $e^+e^-$: electron-positron pair. $\gamma$: bremsstrahlung photon. Decay $\pi^{-,0,+}$: pion issued upon decay. Decay $K^{-,0,+}$: kaon issued upon decay. Decay $e^{-,+}$: electron or positron issued upon decay. PN hads: hadrons created by photonuclear interaction. See text for details.}
    \label{fig:numbers_tau}
\end{figure}

In order to estimate the size of the contribution of secondary showers, we discuss the number of showers above \SI{1}{PeV} initiated by a secondary muon or tau as derived from the lepton propagation code PROPOSAL \cite{proposal}.
We have propagated a few thousand taus (positive and negative randomly mixed) uniformly distributed in
log-spaced energy bins from \SI{1e15}{eV} to \SI{1e20}{eV}. We have gathered all the shower-inducing secondaries above \SI{1}{PeV} and
calculated their numbers. We show in Fig.~\ref{fig:numbers_tau} the average number of showers produced 
by a tau as a function of energy and classified by the type of secondary particle that initiates the shower. Around $\sim\SI{5}{PeV}$, the tau secondaries produce
one shower on average, and the number of showers increases with tau energy, reaching several thousands at \SI{1e20}{eV}. The different primaries are electrons and positrons produced upon decay (decay e), bremsstrahlung photons ($\gamma$), pions produced upon decay ($\pi^{-,0,+}$), kaons produced upon decay (K$^{-,0,+}$), a set of several hadrons produced after decay (decay hads), electron-positron pairs ($e^+e^-$), and a set of hadrons produced after a photonuclear interaction (PN hads). If more than one hadron is produced upon decay, it is counted as a single shower produced by decay hadrons. Pions and electrons produced during decay are dominant below \SI{\sim 20}{PeV}, and as energy increases, decay hadrons become more prevalent. Around \SI{40}{PeV}, photonuclear hadrons and pair production are the dominant processes, and above \SI{1}{EeV} electron-positron pairs become the most numerous primaries by more than one order of magnitude, although they carry less energy on average than photonuclear hadrons. The relative numbers can be found in Fig.~\ref{fig:numbers_tau}, bottom. The rest of possible channels, such as ionized electrons, have lower numbers than the scale of the graph and can be ignored for our purposes.

\begin{figure}
    \centering
    \includegraphics[width=0.48\textwidth]{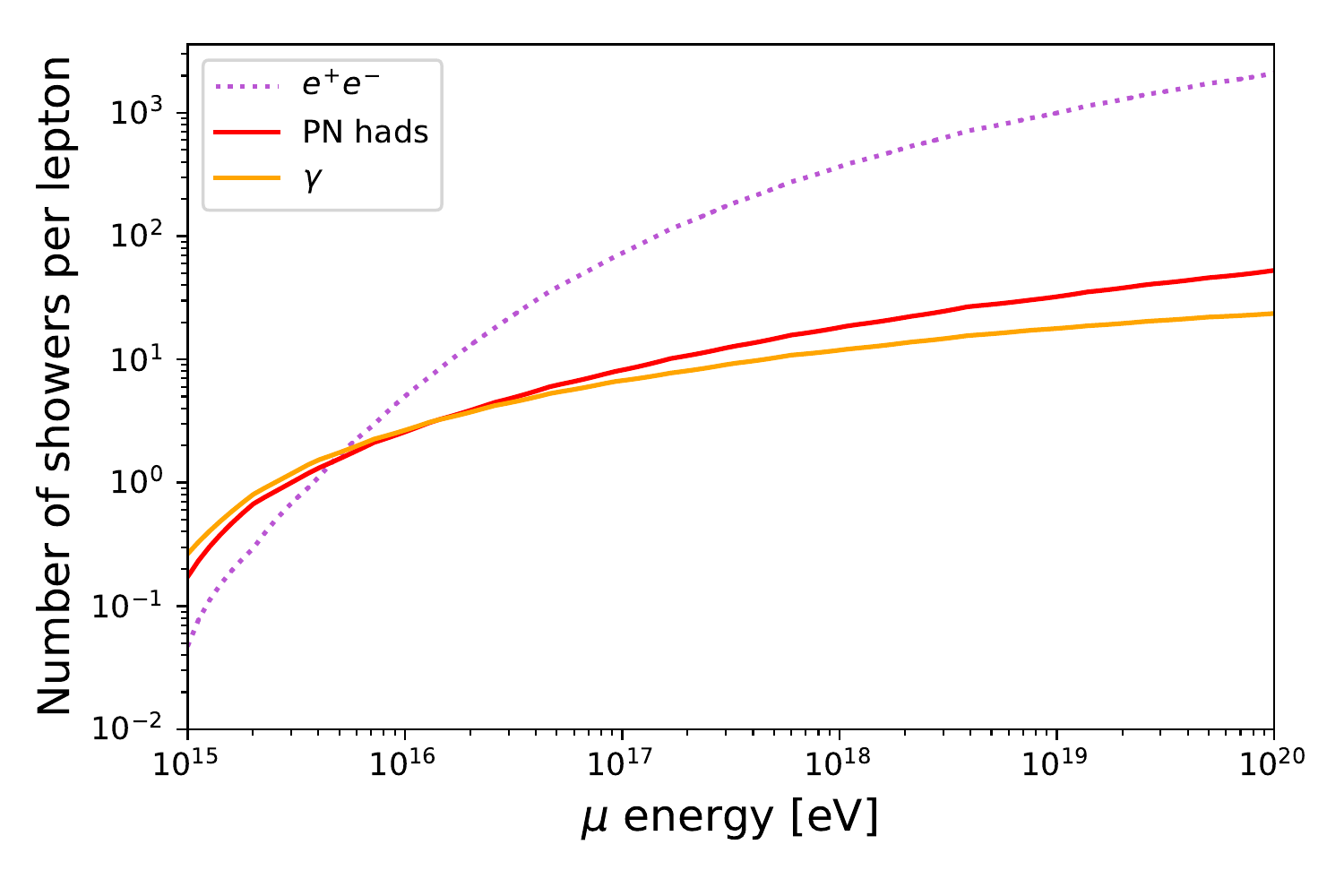}
    \includegraphics[width=0.48\textwidth]{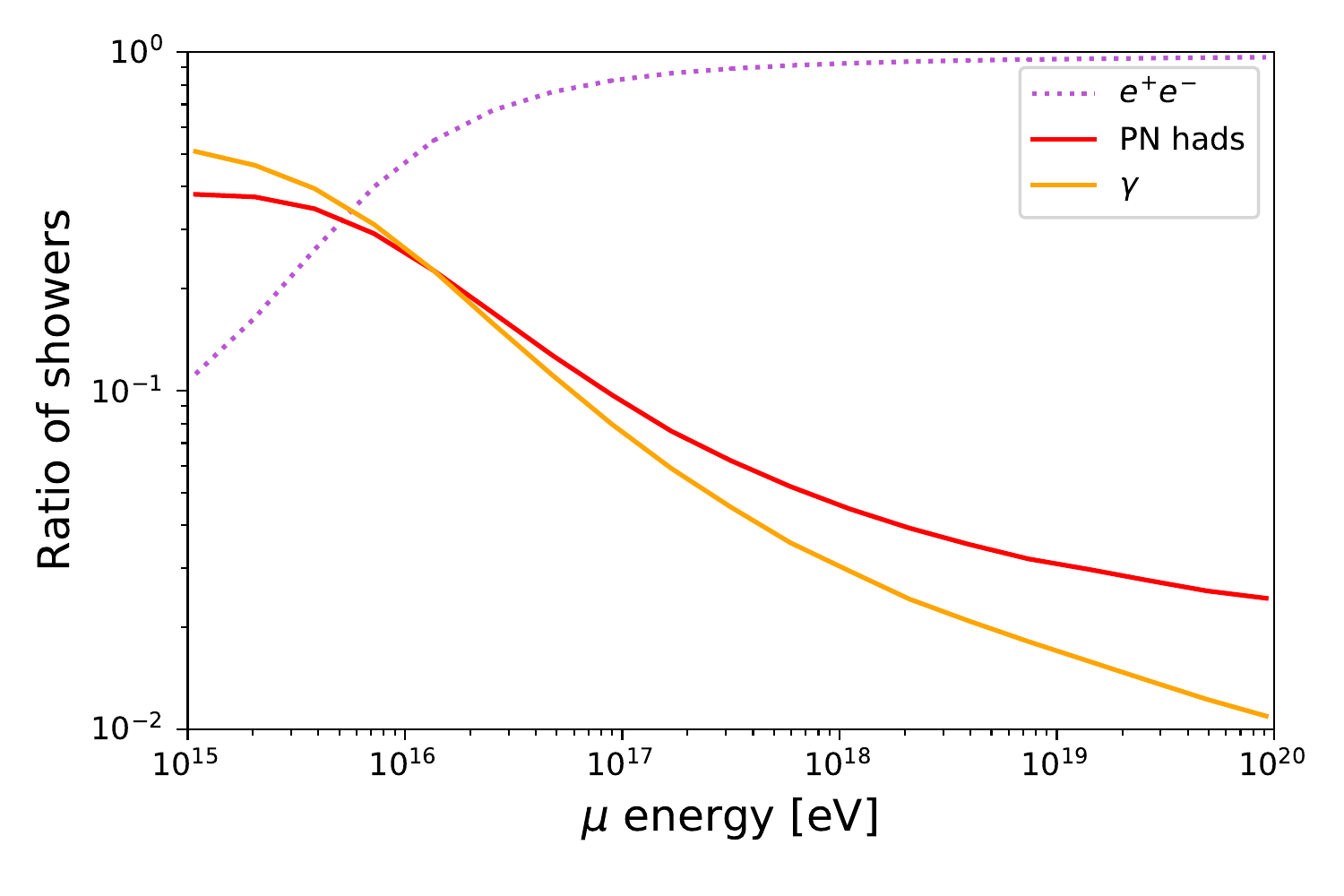}
    \caption{Top: Average number of ($>\SI{1}{PeV}$) showers produced by a muon as a function of the initial energy, classified by primary type. Bottom: Ratio of average number of showers per primary type over total number of particle showers for a muon lepton. The shower primaries in the legend are noted as follows. $e^+e^-$: electron-positron pair. $\gamma$: bremsstrahlung photon. PN hads: hadrons created by photonuclear interaction. See text for details.}
    \label{fig:numbers_mu}
\end{figure}

In Fig.~\ref{fig:numbers_mu}, we show the same study for muons instead of taus. The relevant channels are bremsstrahlung, pair production and photonuclear hadrons, since the probability
of these processes is much larger than any other in the case of muons.

\begin{figure}
    \centering
    \includegraphics[width=0.48\textwidth]{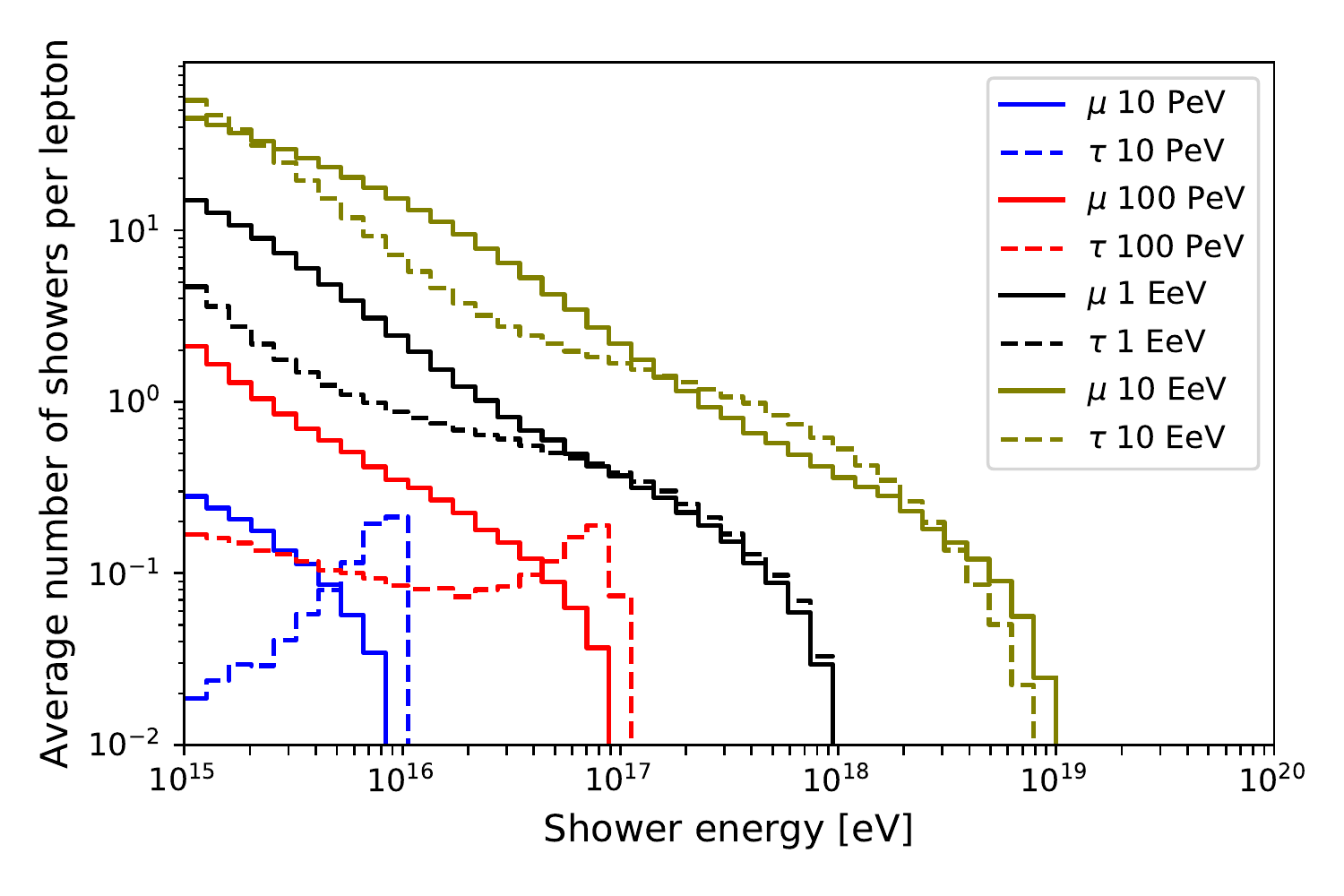}
    \caption{Histogram of the average number of secondary showers per lepton as a function of shower energy. 
    The initial lepton energies are \SI{10}{PeV},
    \SI{100}{PeV}, \SI{1}{EeV}, and \SI{10}{EeV}. Dashed lines represent taus and solid lines represent muons.}
    \label{fig:shower_spectra}
\end{figure}

Figs.~\ref{fig:numbers_tau} and~\ref{fig:numbers_mu} hint at a non-negligible number of radio-detectable, secondary showers stemming from a CC neutrino interaction. We can study the spectrum of these secondary showers for a fixed initial lepton energy. Let us take taus and muons with initial energies of \SI{10}{PeV}, \SI{100}{PeV}, \SI{1}{EeV}, and \SI{10}{EeV} and propagate them to calculate a histogram with the average number of showers per lepton and shower energy bin. The results are shown in
Fig.~\ref{fig:shower_spectra}, where it can be seen that taus having an initial energy around and below \SI{\sim 10}{PeV} radiate more energetic secondary showers than muons of the same initial energy. 
However, when the initial energy is above \SI{\sim 1}{EeV}, muons radiate more showers than taus of the same energy across almost all the shower energy spectrum.
The overall effect we expect is that there should be more detected tau-induced secondary showers for an initial tau neutrino energy around the tens of PeV than detected muon-induced secondary showers after from muon neutrinos of the same energy. If the initial neutrino has an energy above \SI{\sim 1}{EeV}, on the other hand, there should be more detected muon-induced secondary showers than detected tau-induced secondary showers.
We will show that this reflects on the effective volumes for each flavor in Sect.~\ref{sec:effective_volumes}.
It is also worth noting that low-energy taus present a peak near their initial energy and that is due to their decay.
Another relevant fact is that
the most energetic showers from muons and taus having an initial energy greater than \SI{100} PeV are created by photonuclear
interactions and bremsstrahlung mostly, while most of the low-energy showers come from pair production.

It thus seems necessary to study both contributions from secondary muons and taus to the neutrino effective volume, as well as the potential background from atmospheric muons in detail. 


\section{Simulation set-up}
\label{sec:simulations}
In order to study the contributions from secondary leptons, the sensitivities of pathfinder arrays to these interactions are studied.
In order to study the contributions from secondary leptons, NuRadioMC has been extended to use the lepton propagation software PROPOSAL \cite{proposal,Dunsch:2018nsc}.
We have performed simulations for four different different configurations: dipoles at \SI{5}{m} of depth, dipoles at \SI{100}{m} of depth, stations triggered by phased arrays at \SI{100}{m} of depth, and ARIANNA-like surface stations.
The aim of simulating the two sets of dipoles at different depths is to assess the depth dependence of the number of detected atmospheric muons for a conceptually simple detector, as it is known that the detected number of neutrinos goes up with increasing depth and it would be instrumental to know if the detected atmospheric muons behave similarly.
The reason of simulating the phased array and the surface stations is to give a more realistic prediction for the background.  The phased array configuration used is similar to the planned RNO-G detector at Summit station in Greenland, which uses a phased dipole array trigger at \SI{100}{m} of depth \cite{Aguilar:2019jay}, while the surface station configuration is close to the proposed ARIANNA-200 array on the Ross Ice shelf which equips each station with log-periodic dipole antennas near the surface \cite{2020ARIANNA-200}.

\subsection{Usage of the lepton propagation code PROPOSAL}

PROPOSAL \cite{proposal,Dunsch:2018nsc} is a Monte Carlo code that calculates the propagation of leptons (i.e.~muons and taus) in a medium. It has been extensively used and tested by the IceCube Collaboration, e.g.~\cite{Aartsen:2015nss,Aartsen:2017bap,Aartsen:2017nmd,Aartsen:2019tjl}. PROPOSAL calculates all the relevant lepton interactions at these energies, namely, ionization, bremsstrahlung, photonuclear interactions, electron pair production, Landau-Pomeranchuck-Migdal effect and Ter-Mikaelian effect, alongside with muon and tau decay.

NuRadioMC uses PROPOSAL when generating input files. If an input muon neutrino interacts via CC interaction and a muon is created, NuRadioMC can ask PROPOSAL to propagate this muon. Analogously, if a tau neutrino undergoes a CC interaction and a tau is created, PROPOSAL propagates this lepton. NuRadioMC then reads the interactions and particles created during the propagation and saves the ones having an energy $> \SI{1}{PeV}$, since the effective volume below this energy is expected to be negligible \cite{NuRadioMC}. The following cases are considered:
\begin{itemize}
    \item If the interaction is ionization, the resulting electron will create an electromagnetic (EM) shower.
    \item If it is bremsstrahlung, the radiated photon will create an EM shower.
    \item For photonuclear interactions, the outgoing hadrons will create a hadronic shower.
    \item An outgoing electron-positron pair will create an EM shower.
    \item If a muon decays, the resulting electron will create an EM shower. The neutrinos are ignored.
    \item If a tau decays, it can decay either in leptonic or hadronic mode. If there is an electron or positron in the products, it creates an EM shower. The muons produced by the decay are propagated, since they can create showers on their own, while the resulting neutrinos are ignored. If the tau decays into a hadronic channel, the outgoing hadrons will induce a hadronic cascade.
\end{itemize}
We note that ignoring regeneration results is an underestimation of the neutrino flux at low energies, but that detailed calculation lies outside the scope of the present paper and has been done previously elsewhere \cite{regeneration,Safa:2019ege}. Its study is, however, in principle possible using PROPOSAL and NuRadioMC combined.

Throughout this work, we have used PROPOSAL version 6.1.1 to create the input files.

\subsection{Detector layout}
\label{sec:detector}

For this study, we have chosen Summit Station in Greenland as detector location. The ice thickness is \SI{\sim 3}{km}, its
refractive index can be derived from \cite{alley_1988,hawley_2008}, and its attenuation length has been reported in \cite{greenland_att}.

We will only consider triggering events to calculate the effective volumes. No analysis efficiency or other criteria like coincidences of multiple antennas are required. 
For simplicity, we will use a single dipole antenna per station at \SI{100}{m} (or \SI{5}{m}) of depth as proxy for a fully phased array trigger made out of dipoles and located at the same depth. An event triggers, if the voltage at the antenna reaches $1.5\sigma$, with $\sigma$ the voltage RMS of thermal noise at \SI{300}{K}. We have chosen $1.5\sigma$ because it is one of the most efficient thresholds that could be experimentally available by using an in-ice phased array for triggering \cite{Allison:2018ynt}.
The used bandwidth spans from 
80 to \SI{500}{MHz}, with a Butterworth high-pass filter of order
2 and a low-pass filter of order 10. 

The choice of the $1.5\sigma$ threshold is justified by it being a proxy for the performance of the phased array \cite{Allison:2018ynt} that can be used in a simple detector, making it less computationally expensive to simulate. The threshold is considered ambitious, but suitable to give a worst-case estimate for the number of background muons. Any other trigger would likely lower the number of triggered muons and secondary leptons, and would change the distributions shown in the present work. However, the relative numbers (e.g. the effective volume from secondary interaction relative to the effective volume of first interactions) should be independent of changes to the trigger threshold to first order. We must note, however, that we are only considering trigger efficiency, and
the distributions will also change with analysis cuts. It is not our aim to
carry out a systematic study of all possible triggers and analysis cuts, as it appears as a lengthy task beyond the scope of the present work.

The signal measured at the given depth in the dipole will determine if a neutrino candidate event triggers the detector, so it is sufficient to estimate the neutrino effective volume. The remaining antennas of the detector station serve the purpose of event reconstruction, which lies beyond the
scope of this work. Each station will be approximated, thus, as a single dipole, and a whole array will be approximated as a set of dipoles at a distance of \SI{1.25}{km} to each other.

We will use different station and trigger layouts, however, when discussing atmospheric backgrounds, such as a dipole close to the surface and a completely simulated phased array. This is to illustrate the effect of the trigger on background. It is not our intention to find the optimal trigger layout.


\section{Atmospheric muons}
\label{sec:atmospheric}

Cosmic-ray showers contain muons that
reach ground level and can radiate in the ice inducing a signal that can trigger an in-ice radio detector, given that the subsequent shower has sufficient energy. Since this radio signal is created by an
either hadronic or electromagnetic particle shower, the signal would look, in principle, identical to the signal created by a
neutrino-induced particle shower. Since we expect a constant atmospheric cosmic-ray-induced muon flux, these muons may constitute an
atmospheric background. In this section, we will calculate the effect of such a flux on an in-ice radio array to assess the need for an
air-shower veto.

\subsection{Atmospheric muon flux}

Given the limited available data for the absolute atmospheric high-energy muon flux, we use the MCEq code (version 1.2.1) for solving cascade equations \cite{mceq_sibyll, mceq}. This
code allows the user to choose from a variety of primary cosmic ray fluxes and solves the cascade equations for the chosen primaries. We will use the global spline fit model for the primary cosmic-ray flux and composition from \cite{GSF} throughout this paper.
The cross sections can be taken from several models
(e.g.\ SYBILL \cite{sibyll_arxiv}, EPOS-LHC \cite{epos-lhc}, and
QGSJet \cite{qgsjet}). The combination of primary cosmic ray flux and composition, and hadronic interaction model
returns the fluxes of every secondary particle at the desired altitude. 

IceCube has directly measured the muon flux up to the PeV scale \cite{Aartsen:2015nss}, i.e. below the energy range we are interested in, and found some inconsistencies in the
data/MC comparisons of the angular distributions that could be attributed to the hadronic models used.

With the Pierre Auger Observatory it has been found that predicted muon fluxes present discrepancies regarding the muon distributions created by ultra-high-energy
cosmic ray showers (CR energy $>10^{18.4}$~eV) at muon energies \SI{>0.3}{GeV}, indicating important biases in the hadronic models \cite{muon_puzzle1,muon_puzzle2}.
One of the goals of the EPOS-LHC model was to reconcile the discrepant results obtained by air-shower arrays, but
the problem has not been completely solved as of now \cite{pierog2019, Cazon:2020zhx}. 
As a consequence, the atmospheric muon flux above PeV energies is not well constrained and the models predict a large range of possible
muon fluxes.

\begin{figure}
    \centering
    \includegraphics[width=0.48\textwidth]{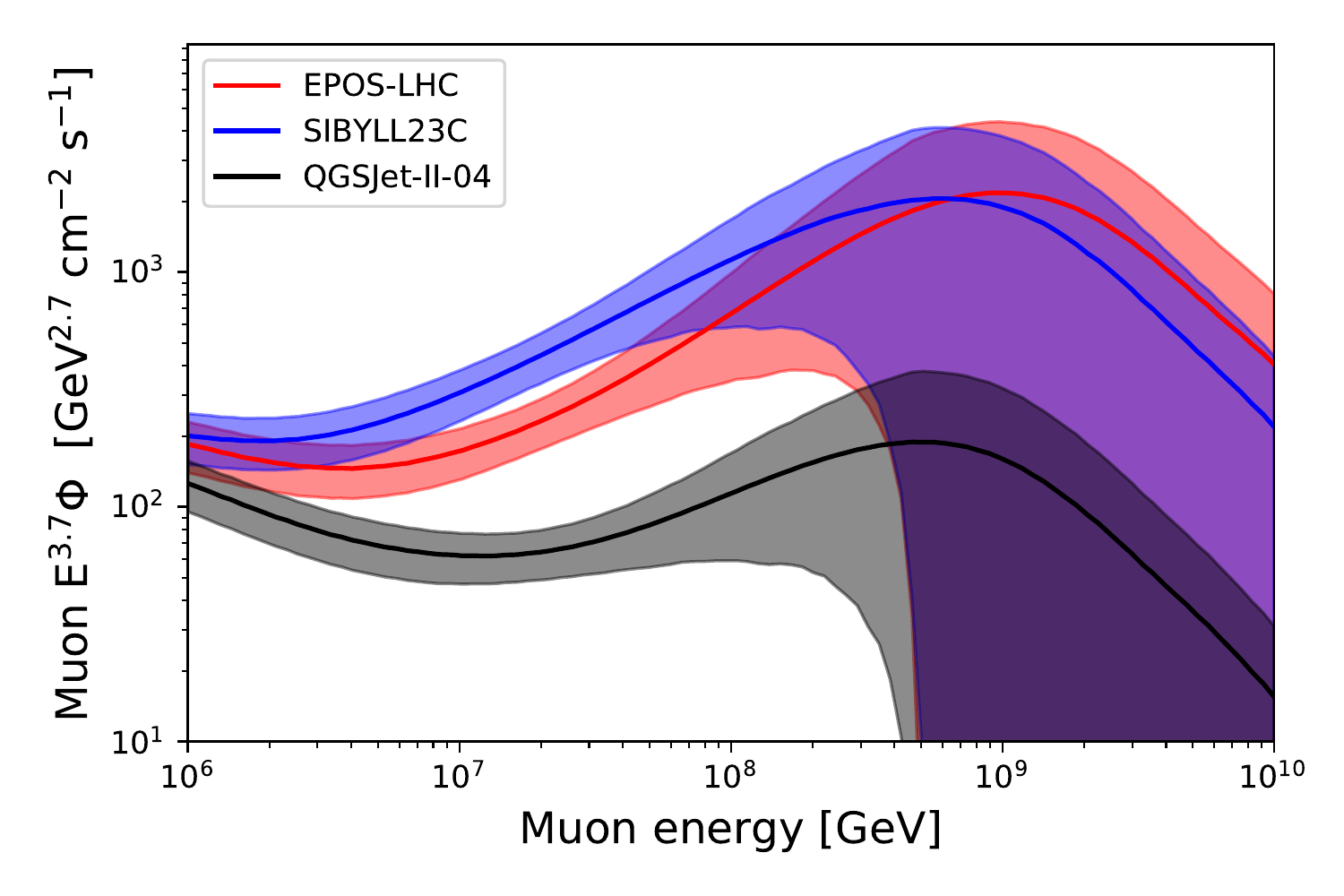}
    \caption{Muon flux rescaled with muon energy to the power of $3.7$ and integrated on the upper hemisphere, given in \SI{}{GeV^{2.7} cm^{-2} s^{-1}} as a function of the muon energy.
    These curves have been obtained using MCEq (version 1.2.1) with the cosmic ray
    model from \cite{GSF} and using four different models:
    EPOS-LHC (red), SYBILL23C (blue), and QGSJet-II-04 (black). The observation altitude is
    \SI{2.7}{km}. Uncertainties due to cosmic ray flux models and hadronic interaction models \cite{state_leptons} are represented by the shaded region (68\% CL). }
    \label{fig:muon_flux}
\end{figure}

We show in Fig.~\ref{fig:muon_flux} a
comparison of the muon flux integrated on the upper hemisphere from three different interaction models: EPOS-LHC, SYBILL23C,
and QGSJet-II-04. We can see that the predictions vary by about an order of magnitude in the relevant energy range. EPOS-LHC yields more muons at higher energies than all the
other models, but at low energies falls short of SIBYLL.
QGSJet is much lower than the rest of the models up to \SI{1e9}{GeV}. The bands represent the combined uncertainty
due to the cosmic ray flux and the hadronic model predictions,
taken from \cite{state_leptons}. Each hadronic model has its own uncertainty induced by the respective modeling of the hadronic
structure and interactions, although these uncertainties are fairly similar across models.
Muon flux uncertainties are dominated by hadronic model uncertainties below \SI{\sim 30}{PeV}, and by cosmic ray flux uncertainties above this energy \cite{state_leptons}.
Above \SI{\sim 0.1}{EeV}, there are no calculations available, so we have
extrapolated using a linear interpolation in log space, where we assume the maximum relative cosmic ray flux uncertainty
to be 100~\%.
We would like to point out that interactions involving charm are important at the energies the in-ice radio experiments can
probe, and SIBYLL 2.3c is the only model that includes charm. We will include EPOS-LHC and QGSJet-II-04 to give an idea about the prediction of different models.
One should keep in mind the large model uncertainties implied
by this figure when discussing the atmospheric muon backgrounds. Still, these models provide the currently best estimate of the atmospheric muon background and will help us establish whether an air-shower veto is mandatory.

\subsection{Signal of atmospheric muons in radio detectors}

For the simulations, we have divided the sky in 8 constant solid angle zenith bands from the zenith to the horizon. For each band, muons were drawn from an
isotropic arrival distribution and a logarithmic energy distribution. We have used 19 log-spaced muon energy bins from \SI{1e15}{eV} to
\SI{1e19}{eV}. The muon initial positions are uniformly distributed on a circle lying on the air-ice planar interface. The radius has been
chosen to be large enough for each zenith angle bin such that the muons generated at the circumference have a negligible trigger probability.
In other words, muons generated outside this circle will not trigger the array and therefore the circle contains all the relevant phase space for the problem.

Once the muon position, energy, and direction have been drawn, we use the PROPOSAL module embedded in NuRadioMC to propagate the muons and 
determine where they create showers above the \SI{1}{PeV} threshold. These showers are then processed by NuRadioMC to calculate the
number of triggers induced by these muons, which can be used to estimate the average effective area for each energy and zenith bin, given by:
\begin{equation}
    \langle A_\textrm{eff} \rangle = 
    \langle A_\textrm{proj} \rangle \frac{N_\textrm{trig}}{N_\mu} =
    A_\textrm{sim} \frac{\cos\theta_1+\cos\theta_2}{2}
    \frac{N_\textrm{trig}}{N_\mu},
    \label{eq:a_eff}
\end{equation}
where $\langle A_\textrm{proj} \rangle$ is the average projected area. This is needed to get a correct effective area normalization and can
be obtained using the area of the simulated circle, $A_\textrm{sim}$, and the edges of the zenith bin, $\theta_{1,2}$. $N_\textrm{trig}$
represents the number of muons that trigger, and $N_\mu$ the total number of input muons of the zenith angle bin.

We must note that Eq.~\ref{eq:a_eff} presents no interaction
probability or mean free path terms because they are already 
taken into account by the muon propagation carried out by PROPOSAL, that is, the interaction probability is not forced as PROPOSAL has
randomly distributed the muon interaction vertices. Moreover, energy losses are calculated by PROPOSAL.

This effective area can be multiplied by an incident muon flux integrated in an energy bin and a zenith band to arrive at the
average number of events triggered by atmospheric muons, which characterizes the influence of the atmospheric background on the detector.
The muon fluxes at the surface have been obtained using MCEq.

Eq.~\ref{eq:a_eff} can be used for an array. However, to save computational time we will simulate a single station
and multiply the results by the number of stations, which we will take equal to $100$. The actual number of triggering
events will be lower, since the stations can have overlapping effective volumes \cite{NuRadioMC}, depending on the distance between stations. This overlap is typically less than 10\% at \SI{e18}{eV} \cite{NuRadioMC} and smaller at lower energies. Still, this calculation is enough to
provide the order of magnitude of the number of muons, and in any case the total uncertainty is much larger than the error induced
by extrapolating the results from a single station.

We have simulated several different configurations. The first one consists of a \SI{100}{m}-deep dipole with a $1.5\sigma$
amplitude threshold, with $\sigma$ being the noise RMS. The second is the same configuration placed at a depth of \SI{5}{m}. 
Appendix~\ref{sec:triggers} discusses how much a realistic trigger and the selection of location for the experiment changes the
muon background.

Muons with energies from \SI{1}{PeV} to \SI{100}{EeV} have been simulated for the aforementioned 8 zenith bands and the
effective area as a function of muon energy has been calculated. The expected number of muon events can be
obtained as a function of muon energy directly, however,
since the muon energy is not directly observable
with radio experiments, it is more desirable to express them as a function of cosmic-ray energy. This is useful because
the acceptance of an air-shower detector is usually given as a function of cosmic-ray energy, which
helps to discuss the possibility of a surface veto.
To that end, we have calculated the muon flux using MCEq for different cosmic-ray energy bins and combined these 
fluxes with the effective areas.

\begin{figure}
    \centering
    \includegraphics[width=0.48\textwidth]{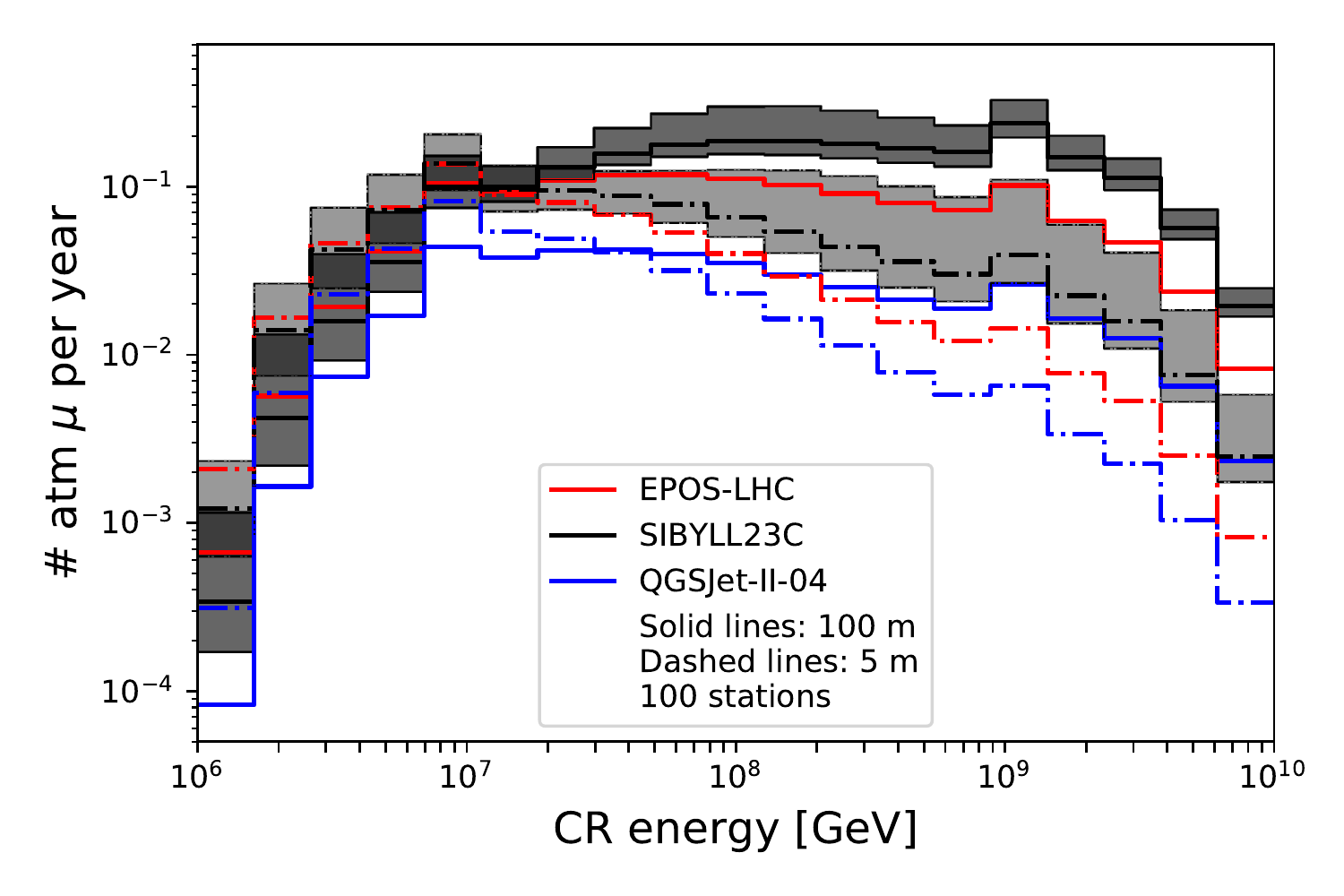}
    \includegraphics[width=0.48\textwidth]{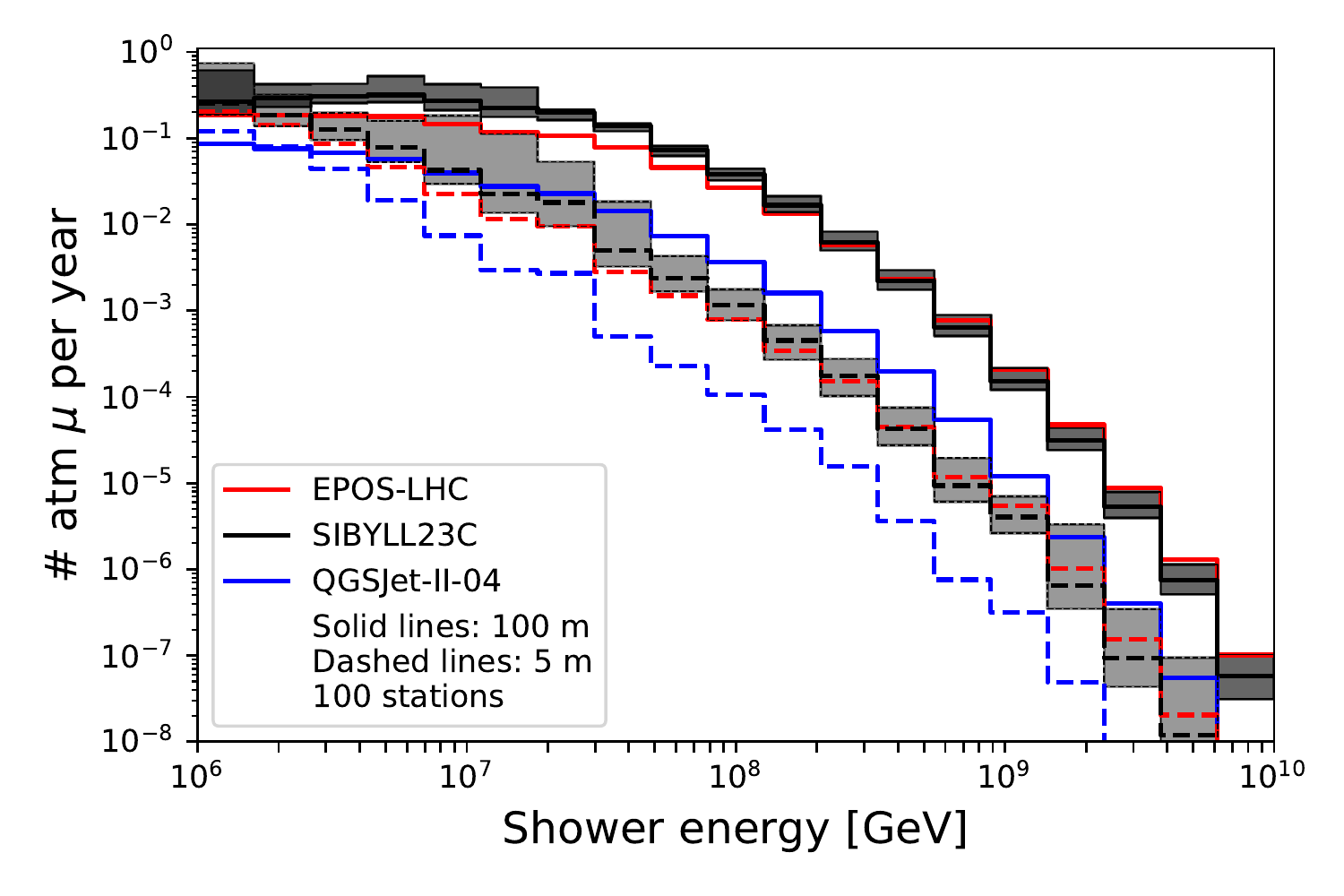}
    \caption{Histograms containing the average number of atmospheric muons detected by a 100-station array of $1.5\sigma$ dipoles at
    \SI{100}{m} (solid lines) and \SI{5}{m} of depth (dashed lines). 
    The shaded bands represent the uncertainties induced by the cosmic-ray flux model, the hadronic model, and the statistical uncertainty of the effective area calculation, for the SIBYLL 2.3C model only.
    Shown are different hadronic models. Top: as a function of
    cosmic ray energy in GeV. Bottom: as a function of the shower energy.}
    \label{fig:muon_numbers}
\end{figure}

We show in Fig.~\ref{fig:muon_numbers}, top, the muon numbers as a function of cosmic-ray energy for dipoles at
\SI{100}{m} and \SI{5}{m} of depth. A deep array sees more muons than a shallow one. The reason is that for shallow
antennas, the paths a ray can take from shower to observer and trigger the detector are fewer than for a deeper array \footnote{Radio signal propagation is approximated by ray optics, for more detail please refer to \cite{NuRadioMC}}, which in turn restricts the
number of allowed vertex positions for the muon to interact at, effectively reducing the effective volume. Another way
to put this is that the volume where muons can interact and trigger the detector is larger when the antennas are deeper
in the ice (this reasoning can also be applied to incident neutrinos or any other incident shower-inducing particle).
We show in Fig.~\ref{fig:muon_numbers}, bottom, the same background numbers as a function of shower energy (the energy
of the shower created by the muon). 
Unlike muon energy,
shower energy is an observable that can be reconstructed using the measured electric field. The distribution peaks below \SI{1e7}{GeV}.

The average total number of detected muons per year for the different dipoles at \SI{5}{m} and \SI{100}{m} can be found in
Table~\ref{tab:muon_numbers}, where they are also classified per hadronic interaction model. For completeness, we also include in Table~\ref{tab:muon_numbers} a column with the results for the realistic phased array (see discussion in Appendix~\ref{sec:triggers}).
For a deep dipole, the yearly numbers range from $0.43$ to $2.19$, whereas for a deep phased
array they vary from $0.048$ to $0.311$. 
The deep dipole detects more muons than the phased array because the $1.5\sigma$ threshold for the dipole is lower than the phased array that is planned to be used for the \hbox{RNO-G} experiment.
There is a large variation in detected muon numbers due to the difference in interaction models.
QGSJet predicts the fewest and SYBILL 2.3c predicts the most. The uncertainty of these models will be one of the dominant systematic uncertainties for a given detector system, together with the cosmic ray flux uncertainty, since the uncertainty due to the detector effective area can always be refined with better simulations.

\begin{table}
\begin{center}
\begin{tabular}{ |c|c|c|c| } 
 \hline
                 & Dipole, \SI{100}{m} & Dipole, \SI{5}{m} & PA \SI{100}{m} \\ \hline \hline
 SIBYLL 2.3C     & 2.19 & 0.943 & 0.311 \\ \hline
 EPOS-LHC        & 1.31 & 0.716 & 0.185 \\  \hline
 QGSJet-II-04    & 0.43 & 0.408 & 0.048 \\ \hline
\end{tabular}
\caption{Average number of detected atmospheric muons per year 
for a 100-station array. Three hadronic models and three antenna configurations
are shown. The dipoles have a $1.5\sigma$ threshold and they are
located at 100 (left column) and \SI{5}{m} of depth (center column). PA stands for a realistic envelope phased array at
\SI{100}{m} of depth. 
The relative uncertainties due to cosmic ray flux, hadronic modeling, and effective area are similar across models. The uncertainty on the detected muon numbers for the SIBYLL 2.3C are $\sim {}^{+1.3}_{-0.6}$ for the \SI{100}{m} dipole,
$\sim {}^{+1.0}_{-0.4}$ for the \SI{5}{m} dipole, and $\sim {}^{+0.4}_{-0.1}$ for the phased array.
See text for details.}
\label{tab:muon_numbers}
\end{center}
\end{table}

\subsection{Consequences for neutrino detection}

Given the numbers in Table \ref{tab:muon_numbers}, muons constitute a non-negligible background, if the expected total neutrino flux in a 100 station array yields of the order of one detected neutrino per year. Since the predicted neutrino fluxes vary by about an order of magnitude, a cosmic-ray veto seems advisable. 

Apart from relying on a cosmic-ray veto, one can ask if there is a way to distinguish atmospheric muons from real neutrino events provided, for instance, a reconstructed vertex position, particle arrival direction and the shower energy. Even though there are large variations in the predictions of the neutrino flux, the neutrino energy spectrum is (in the majority of models) significantly harder than the atmospheric muon spectrum that drops with approximately $E_\mu^{-3.7}$. Many cosmogenic neutrino flux models peak around \SI{e9}{GeV} (see e.g.~\cite{van_Vliet_2019}) and the astrophysical neutrino flux measured by IceCube develops with $E^{-2.9} - E^{-2.2}$  \cite{Haack:2017dxi,Kopper:2017zzm}. Hence, in general the signal-to-background ratio will improve with increasing energy. In simplified studies using the above mentioned fluxes, we find that the background-to-signal ratio is likely to improve six orders of magnitude from the energy threshold of the
array of \SI{e7}{GeV} to an energy of \SI{e10}{GeV}. Thus, it is to be anticipated that the shower energy will be an important tool to obtain a signal region with reduced background contamination similar to the analysis of optical neutrino detectors \cite{PhysRevLett.113.101101}. 

As a practical tool, we note that the energy dependent number of background events can be calculated from Fig.~\ref{fig:muon_numbers} (bottom) where the uncertainty of the reconstructed shower energy determines how each bin should be smeared to account for experimental uncertainties and translate the plot from true shower energy to measured shower energy. We also note that for neutrinos, the shower energy is relatively close to the neutrino energy \cite{DnR2019}, which allows for a quick first-order estimation of the background as function of neutrino energy and energy resolution. 

\begin{figure}
    \centering
    \includegraphics[width=0.48\textwidth]{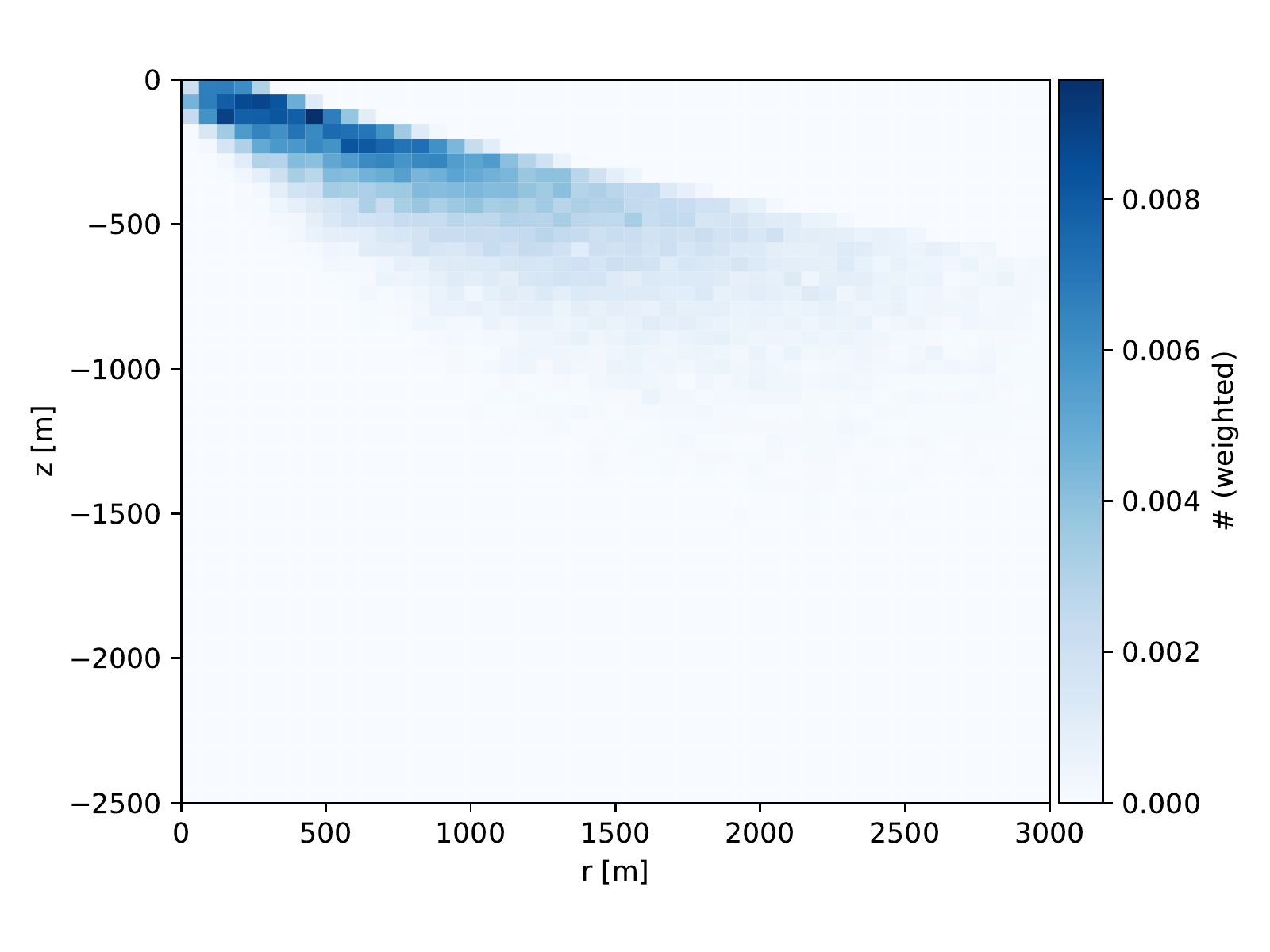}
    \includegraphics[width=0.48\textwidth]{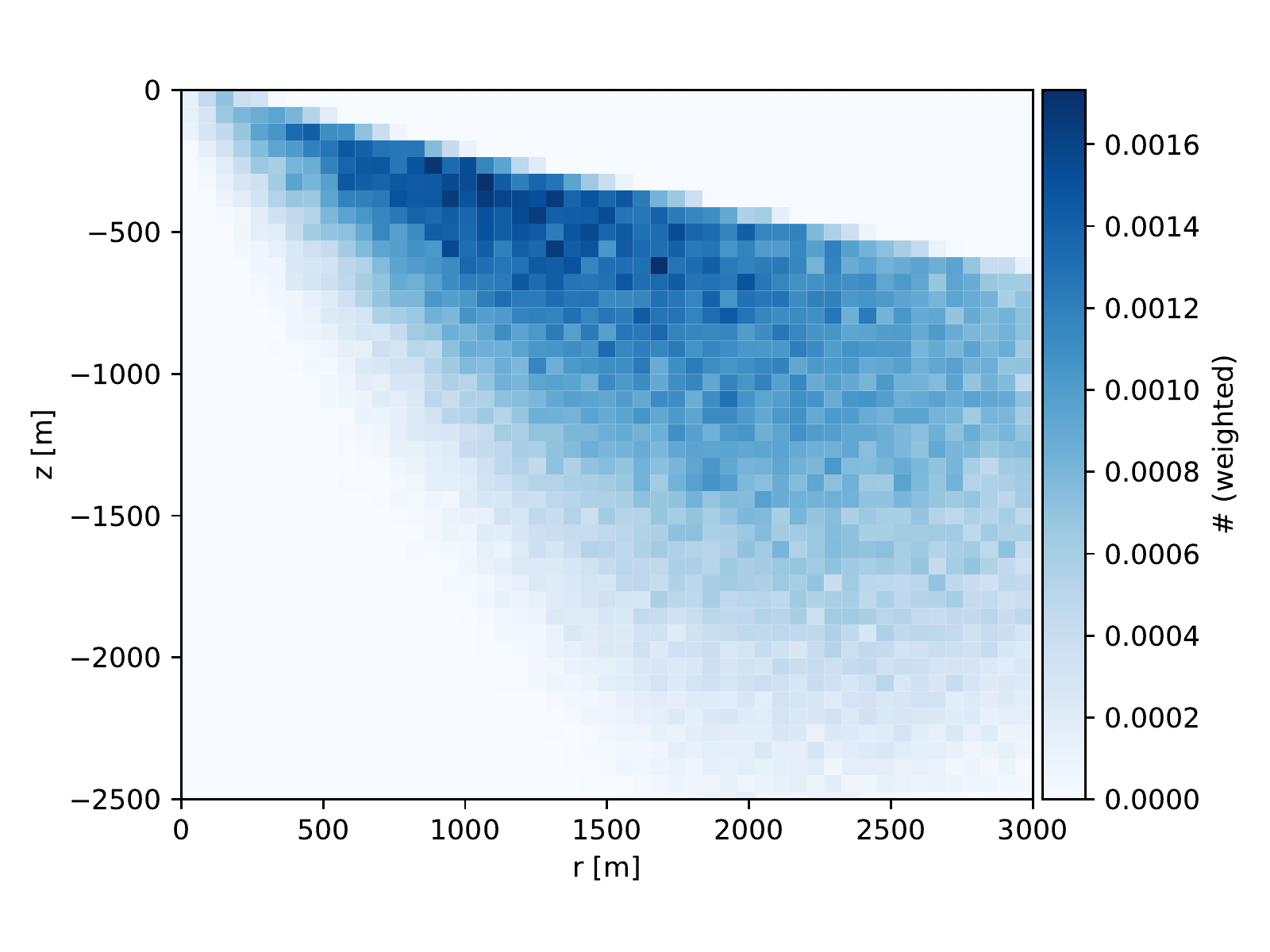}
    \caption{Top: 2D triggering muon vertex distribution as a function of radial and vertical distances to antenna,
    for a $1.5\sigma$ dipole at \SI{100}{m}. Muon energies lie between \SI{300}{PeV} and \SI{6100}{PeV}.
    Bottom: same as top, but for triggering events induced by neutrino first interactions. Neutrino energies lie between \SI{1200}{PeV} and \SI{2300}{PeV}.}
    \label{fig:vertices}
\end{figure}

We also studied if the particle arrival direction or the vertex position, which are accessible from the measurement, can be used to distinguish neutrinos from muons. 
We show in Fig.~\ref{fig:vertices}, the vertices where the triggering muons radiate a shower and the vertices where the triggering neutrinos interact. The distributions are similar as they are dominated by allowed ray-tracing paths, so there is no easy way of telling muons apart from neutrinos simply by the reconstructed vertex location. However, it is also visible that the two distributions are not identical, so one may be able to distinguish events on a statistical basis as neutrinos interact deeper in the ice and muon events tend to be shallower and closer to the detector.

\begin{figure}
    \centering
    \includegraphics[width=0.48\textwidth]{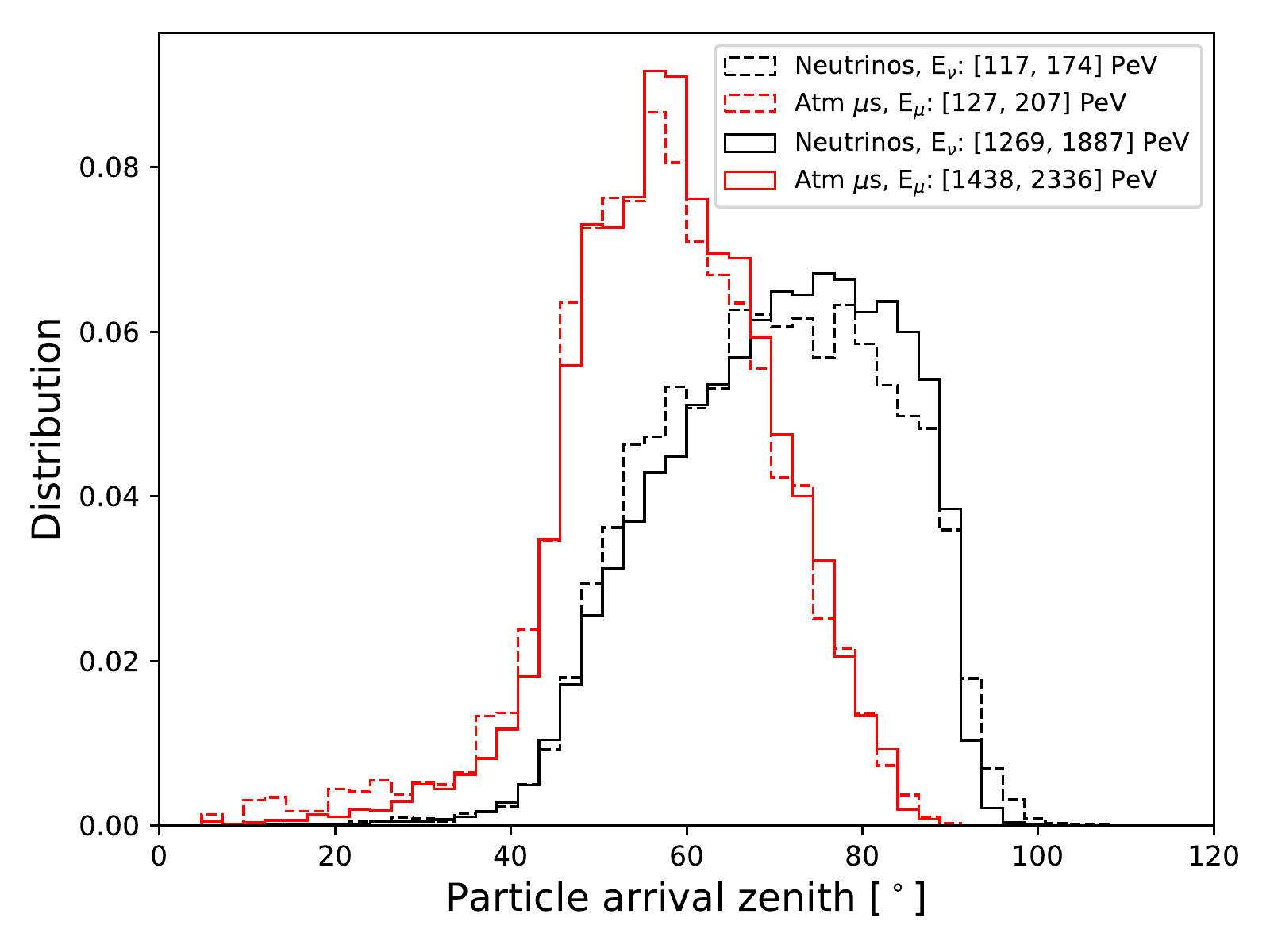}
    \caption{Neutrino (black) and atmospheric muon (red) zenith arrival directions. Solid and dashed lines indicate two
    different energy bins with ranges stated on the legend. For these two bins, the energy dependence seems to be weak. All the shown distributions are normalized to 1.}
    \label{fig:zeniths}
\end{figure}

Finally, Fig.~\ref{fig:zeniths} shows the incoming particle zenith distributions for atmospheric muons and neutrinos for two energy bins. Muon background
starts to trigger around \ang{40}, at about the same elevation where neutrino triggers start to be observed.
However, near the horizon, the number of muons decreases dramatically due to the projection effect, that is, the muon flux projected on a unit surface area drops with the cosine of the zenith angle, and because many inclined muons get absorbed in the ice before reaching the detector. 
Neutrinos, on the other hand, do not suffer from this problem for above and near the horizon because of the much smaller cross section. Below the horizon, and above tens of PeV, the Earth is opaque to
neutrinos, so few neutrinos are seen from below the horizon.
Judging from Fig.~\ref{fig:zeniths}, muons peak around \ang{50}, and neutrinos are more likely to come from close to
the horizon, again hinting at a probabilistic method of distinguishing neutrinos from atmospheric muons based on their arrival direction, even without an air shower veto. 

The vertex distribution (Fig.~\ref{fig:vertices}) can also be combined with the zenith distributions (Fig.~\ref{fig:zeniths}) to create cuts in vertex position and incoming zenith angle that may allow the removal of a fraction of atmospheric muon background, given a good reconstruction of both parameters and at the cost of a reduced neutrino efficiency.

We must stress that the results of this calculation depend critically on the assumptions made for cosmic ray flux and composition, as these determine the total number of muons that reach ground level and their spectrum. As implied by the uncertainties displayed in Fig.~\ref{fig:muon_flux}, the flux can vary orders of magnitude depending on the actual cosmic ray flux and composition (as well as hadronic modeling), which means that the number of detected background atmospheric muons is subject to the same fluctuations.

In general, the predicted astrophysical neutrino fluxes are uncertain. There are imaginable scenarios (e.g. \cite{Heinze:2019jou}) in which the neutrino flux creates less than 1 detectable neutrino at relatively low energy, so suppressing the background is imperative. Fortunately, the most probable energy of the parent cosmic rays that create the relevant muons lies above the threshold for air
shower detection ($\sim 10^{16.5}$~eV) using a sparse radio array at the surface \cite{Huege:2016veh,Barwick:2016mxm}.  In addition, the arrival zenith angles are typically more inclined than \ang{40}, which improves the air-shower detection efficiency of radio arrays. Thus, a surface array of radio detectors is a promising candidate to detect cosmic rays and function as veto, and it can be easily integrated in in-ice radio arrays, as it is based on the same technology, requiring only additional antennas at the surface. More studies beyond the scope of this paper are needed, to estimate the veto efficiency and the reduction rate of the background. 


\section{Muons and taus from neutrino interactions in-ice}
\label{sec:effective_volumes}
Having studied the atmospheric muon background, we now turn to the signals of the secondary showers produced by muons and taus after a CC neutrino interaction has taken place in the ice. For this study, we will assume an isotropic neutrino flux from all directions
arriving at an upright ice cylinder, which we will take as our simulation volume. 

One way of quantifying the
influence the sensitivity of our detector to this isotropic flux is to define an effective volume as follows:
\begin{equation}
    V_\textrm{eff} = \frac{V_\textrm{sim}}{N_\textrm{events}} \sum_{i}^{\textrm{triggered}}
    \omega_i(\theta_i,E_i),
    \label{eq:v_eff}
\end{equation}
where $V_\textrm{sim}$ is the simulated volume, the sum is carried on the \emph{triggering} events, while $N_\textrm{events}$ is the
total number of simulated events, and each $\omega_i$
represents a weight equal to the survival probability of the neutrino going through the Earth and reaching the simulated volume, which is a
function of the incoming direction and the neutrino energy.
These probabilities are calculated analytically using a model based on a spherical Earth divided in three concentric layers of different density and the
cross sections from \cite{Connolly2011}.
Eq.~\eqref{eq:v_eff} is calculated from a Monte Carlo simulation wherein the
interactions are fixed, that is, we consider that all the neutrinos included in the simulation interact. From Eq.~\eqref{eq:v_eff},
and using that the interaction volume dimensions are small with respect to the interaction length, the probability of interaction can be approximated as the ratio of the distance the particle traverses inside the detector volume over the interaction length. If the interaction length is not direction-dependent 
and the simulated volume is large enough
so that increasing it does not change the effective volume calculation, the effective area can be calculated dividing Eq.~\eqref{eq:v_eff} by
the interaction length. The interaction length is direction-dependent if the
detector is sensitive to neutrinos that interact in another medium, e.g. rock, which may create a shower that reaches the ice and triggers our detector. In general, for any detector that sees events from different media as a function of incoming direction, the interaction length is direction-dependent.
Eq.~\eqref{eq:v_eff} for calculating the effective volume is made of a sum on the triggering events, but the effective volume can also be restricted to specific types of interactions.
We can define effective volumes for different types of physical signals: for primary neutrino interactions, interactions from secondary
particles, tau decays, primary \emph{and} tau decay, muon radiative losses, etc, using Eq.~\eqref{eq:v_eff} but including in the sum only a specific type of interaction or channel. 
Since the effective volume is proportional to the number of detected events from that specific type of interaction, we can use
effective volume figures to directly compare the efficiency of our detector to multiple channels.

We must note that when calculating the effective volume with secondary interactions included using Eq.~\eqref{eq:v_eff}
we will consider NC and CC interactions to obtain the total effective volume for each neutrino flavor. The secondary
interactions will only be present for CC neutrino interaction.

\begin{figure*}
    \centering
    \includegraphics[width=0.6\textwidth]{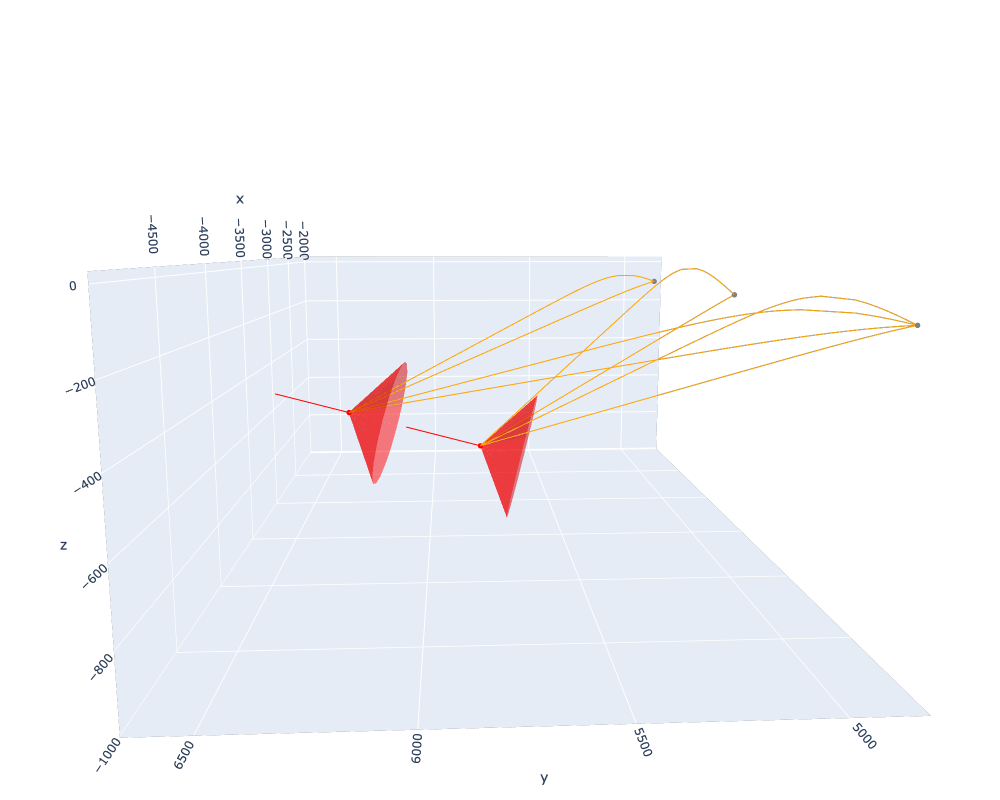}
    \caption{Three dimensional plot of a double-cascade event. The first shower is caused by a \SI{1.84}{EeV} neutrino. 
    The resulting tau travels for several hundred meters and creates a hadronic shower via photonuclear interaction.
    The red line represents the particles trajectories, while the red cones indicate the Cherenkov cones for both showers.
    The three triggered stations are represented by grey dots, and the yellow lines are the paths followed by the waves
    (direct and refracted) that arrive at the stations. The axes units are meters.}
    \label{fig:3d_plot}
\end{figure*}

\subsection{Example of a multiple signature}

The geometry at which double signatures are expected from tau or muon neutrinos is not straightforward to visualize, even for someone familiar with the radio detection of neutrinos. The allowed range of vertex positions is dominated by the allowed ray-paths in the ice and the bending near the surface, as well as the Cherenkov angle with respect to the shower axis.
We have used the event browser from NuRadioReco \cite{NuRadioReco2019} to illustrate what a typical double-cascade event would look like at energies of \SI{\sim 1}{EeV}. Taking for example the
neutrino flux from \cite{van_Vliet_2019} with 10\% protons, we expect the maximum number of neutrinos to trigger around
EeV energies.
The event geometry can be found in Fig.~\ref{fig:3d_plot}, where a tau neutrino
creates a tau that interacts via photonuclear interaction several hundreds of meters after being created and induces a
hadronic shower. The neutrino-induced shower is seen by two of the three stations represented by the grey dots, 
while the tau-induced shower is also seen by two of the three stations. 
Each individual ray in Fig.~\ref{fig:3d_plot} indicates the path and the viewing angle that caused a trigger.

\subsection{Contribution to tau neutrino effective volumes}

\begin{figure}
    \centering
    \includegraphics[width=0.48\textwidth]{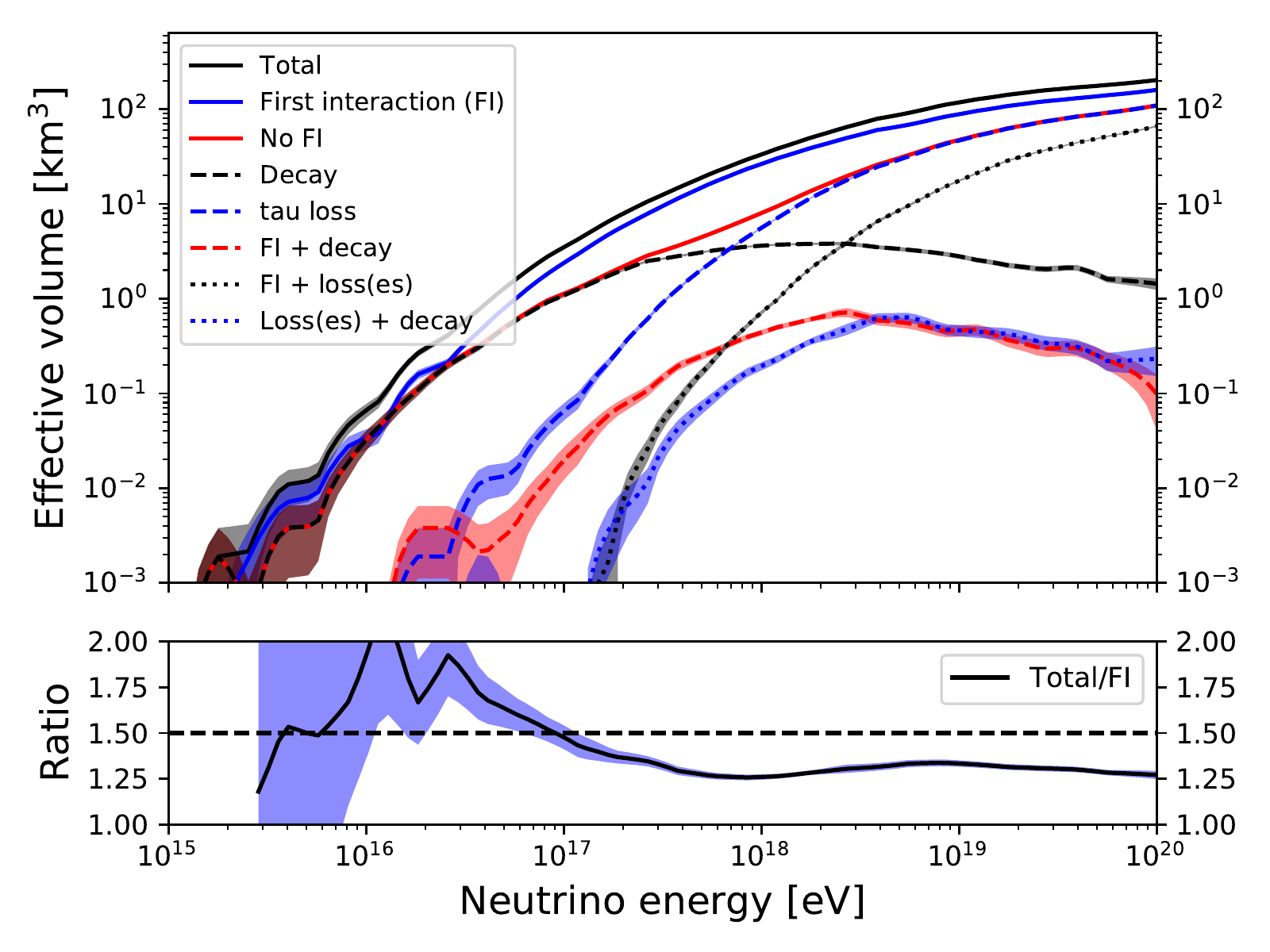}
    \caption{Top panel: tau neutrino effective volumes for a $10\times 10$ square array of \SI{100}{m} deep dipoles, with $1.5\sigma$ threshold.
    The bands represent the uncertainty assuming a Poisson distribution.
    There are different types of effective volume depicted in the figure. The events that have been triggered at least by the shower induced by the first neutrino interaction are represented by the 'First interaction' (FI) curve.
    The events not triggered by the first neutrino interactions but by secondary interactions constitute the 'No FI' curve. 
    The 'decay' curve represents events triggered by the products of the tau decay.
    The events triggered by the tau stochastic losses during propagation result in the effective volume denoted by 'tau loss'.
    The 'FI+decay' curve represents the effective volume from events triggered by the neutrino first interaction \emph{and} the tau decay.
    The curve noted as 'FI+losses' is calculated using events triggered by the neutrino first interaction \emph{and} at least a stochastic energy loss, while the 'Loss(es)$+$decay' shower is the effective volume from events triggered by one or several tau stochastic losses \emph{and} the tau decay. 
    The 'Total' curve contains the total effective
    volume.
    volume. Note that, since the effective volumes are not mutually exclusive, the total curve is not the sum of all the others. Bottom panel: ratio of the total effective volume over the first interaction effective volume.}
    \label{fig:tau_volumes_1}
\end{figure}

\begin{figure}
    \centering
    \includegraphics[width=0.48\textwidth]{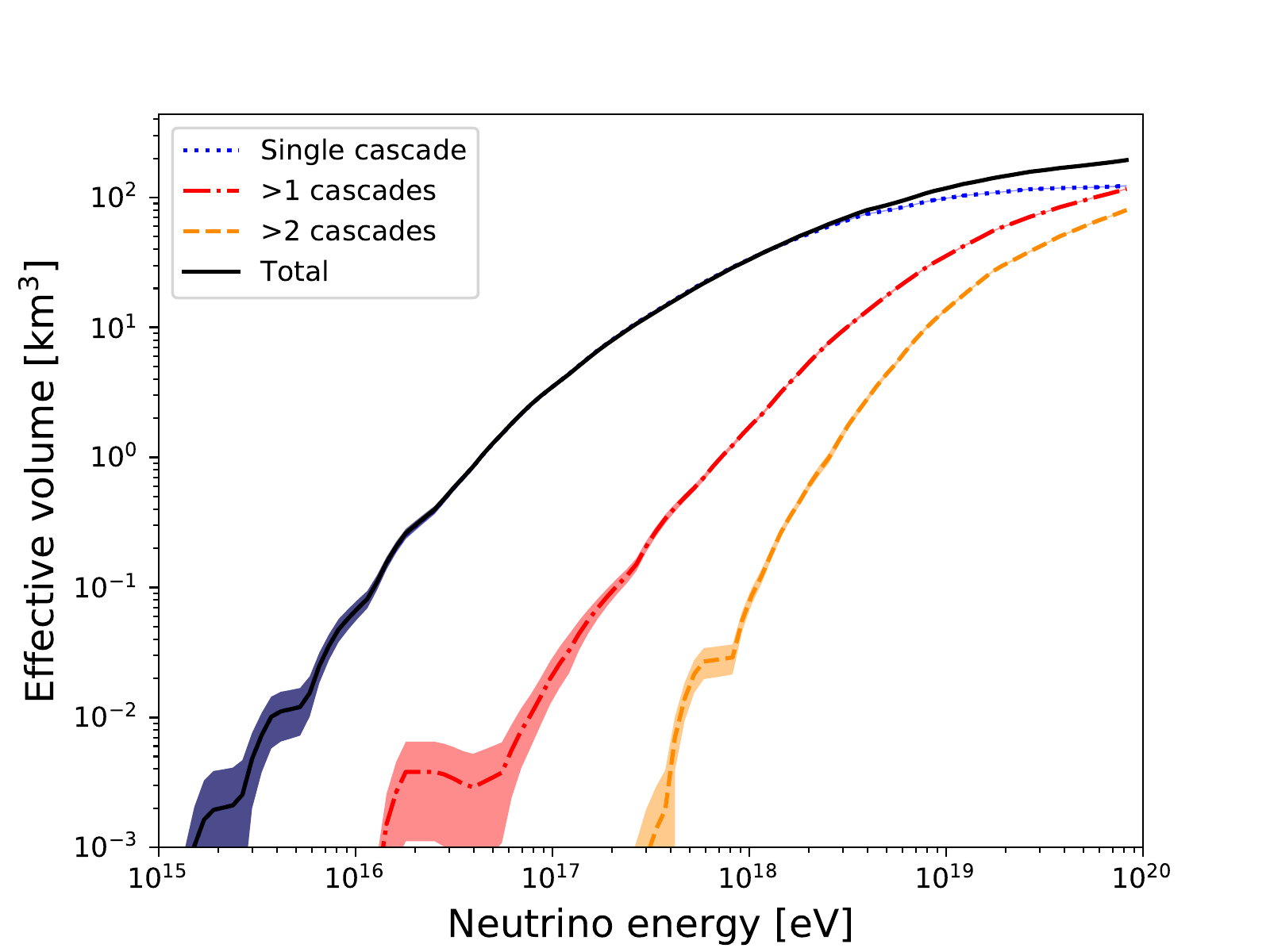}
    \caption{
    Same as Figure \ref{fig:tau_volumes_1}, but with different types of effective volumes: single, events triggered by one interaction (\textit{cascade}) only; multiple cascades, event triggered by more than one interaction; and $>2$ cascades,
    event triggered by more than two interactions. The total effective volume is also shown.}
    \label{fig:tau_volumes_2}
\end{figure}

We have simulated the effective volumes for tau neutrinos detected with a $10\times 10$ array of \SI{100}{m} deep
dipoles, with a spacing of \SI{1.25}{km}. Tau neutrino signals have been simulated with NuRadioMC, without adding noise, using the triggering conditions as discussed in Sec.~\ref{sec:detector}.

Since the tau range grows with energy and can go up to tens of kilometers 
(see Fig.~4 in \cite{NuRadioMC}), we need to choose a simulation volume large enough to contain the interactions along the tau or muon
track as well as its decay. We choose a vertical cylinder with \SI{5}{km} of height and a radius equal to the $95\%$ confidence
interval for the tau range for the maximum energy in a given bin, but using a minimal radius of \SI{11}{km}. The particles are allowed to interact and decay everywhere inside this cylinder,
and then only interactions happening inside a fiducial cylinder of \SI{11}{km} of radius and \SI{3}{km} of height are used
for the calculation of signals and triggering.
The muon range is smaller than the tau range, so we can use the same volume as for taus without losing events. The number of incident
neutrinos has to be increased in the same way as the simulation volume so as to have the same ratio of detected events to interactions
in our fiducial volume, where their signals are detectable. A standard NuRadioMC simulation including ray tracing and using
the parameterization \emph{Alvarez2009} \cite{Alvarez2009} is used and the triggered events are stored. The different types of effective volumes
are obtained by applying Eq.~\eqref{eq:v_eff} to these events.
The Alvarez2009 parameterization contains a simplified way of dealing with the LPM effect that works as follows. 
Electromagnetic showers are randomly stretched following a distribution that has been computed from full LPM simulations
\cite{Alvarez2009}. This approach yields good results for the amplitude of the shower emission, which in turn gives sensible
results for effective volumes. Although actual waveforms are not faithfully reproduced, this is not crucial for the
purposes of the present study. In the case of electron neutrino CC interactions, the Alvarez2009 parameterization is not
accurate when the electromagnetic and hadronic cascade have comparable amplitudes, because it assumes that both showers
are completely coherent, but we must note that the most likely case is that the electromagnetic shower dominates the
amplitude. Since electron neutrinos are not the main objective of this paper, but muon and tau neutrinos, as well
as atmospheric muons, we refer to discussions elsewhere for further detail of the treatment of CC interactions of electron neutrinos \cite{alvarez_lpm,ARZ}. 

The effective volumes from tau neutrinos for different types of triggering interactions can be found in Fig.~\ref{fig:tau_volumes_1}. We can see
that the effective volume for the first interaction produced by the neutrino is dominant, as one would expect given that for all downgoing events, the tau
is not likely to decay in the ice but rather in bedrock. At low energies, the decay channel is the second-dominant interaction channel, but as energy increases, it begins to become less relevant, as only neutrinos
coming in really close to the horizon can stay in the fiducial ice volume before decaying at these energies. 
Stochastic losses (tau loss), on the
other hand, continue to grow with energy and their effective volume increases with energy as well. The multiple-cascade effective volumes
for first interaction and decay, first interaction and loss, and losses and decay are also shown. On the bottom panel the ratio
of the total volume to first interaction volume can be found, representing a 50\% increase around \SI{100}{PeV}, and being
$\sim 25-30$\% above one EeV.

In Fig.~\ref{fig:tau_volumes_2}, the tau neutrino effective volumes for single cascades, multiple cascades (more than 1) and more
than 2 cascades are shown. The multiple-cascade events are more than one order of magnitude below the single-cascade events up to \SI{\sim 10}{EeV}.
At really high energies, the probability of having a multiple-cascade event is not negligible.
Most of these multiple-cascade events are triggered
by either the first interaction and stochastic losses, or by 
two or more stochastic losses along the tau trajectory, as we can see in Fig.~\ref{fig:tau_volumes_1} that decays do not contribute significantly. We can see as well that events with more than two cascades are actually possible to detect given a reasonably large experiment.

\subsection{Contribution to muon neutrino effective volumes}

\begin{figure}
    \centering
    \includegraphics[width=0.48\textwidth]{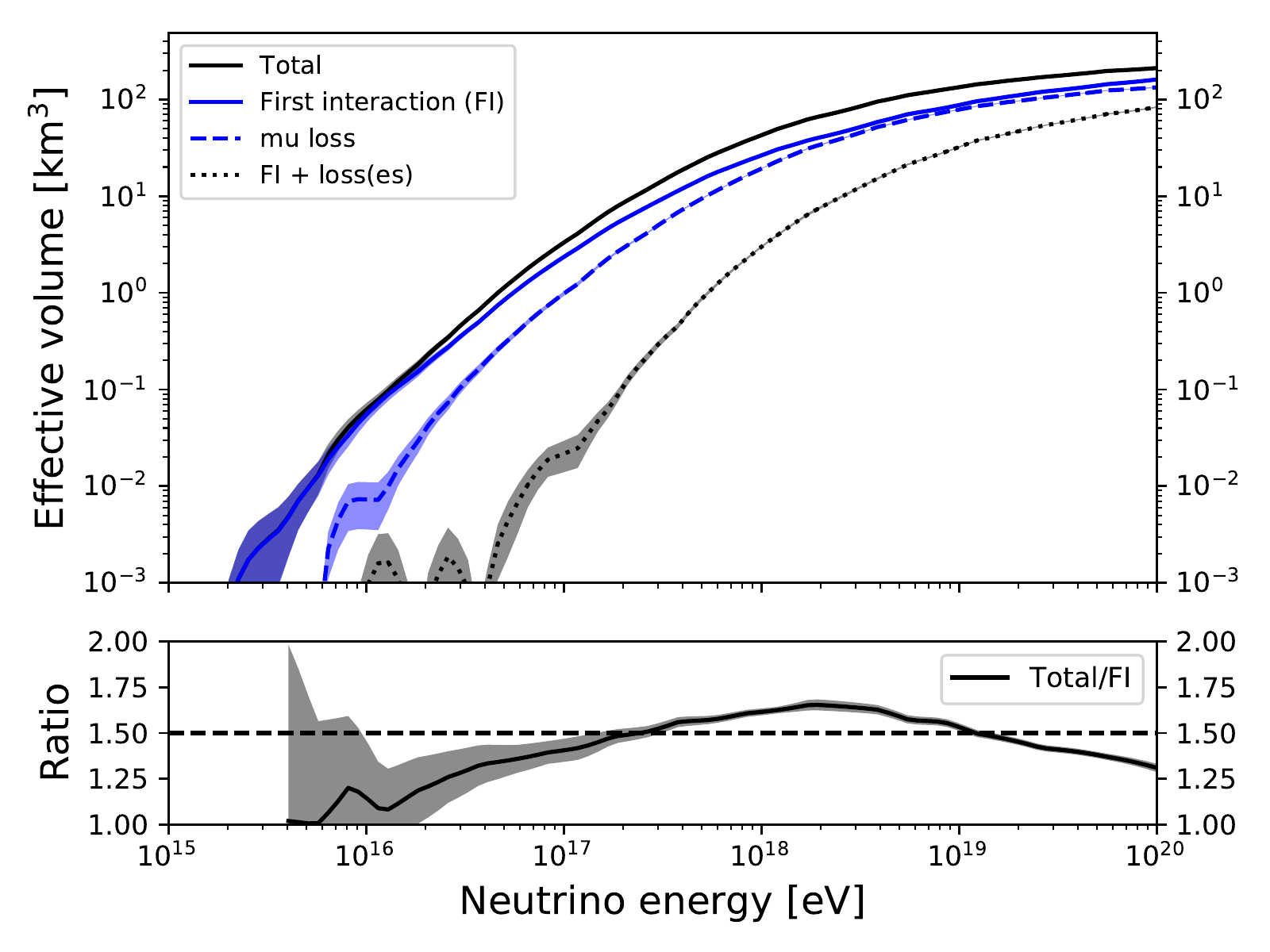}
    \caption{Top panel: muon neutrino effective volumes for an $10\times 10$ square array of \SI{100}{m} deep dipoles, with $1.5\sigma$ threshold.
    The bands represent the uncertainty assuming a Poisson distribution.
    There are different types of effective volume depicted in the figure. The events that have been triggered at least by the shower induced by the neutrino interaction are represented by the 'First interaction' (FI) curve.
    The events triggered by the muon stochastic losses during propagation result in the effective volume denoted by 'mu loss'. The curve noted as 'FI+losses' is calculated using events triggered by the neutrino first interaction \emph{and} at least a stochastic energy loss. The 'Total' curve contains the total effective
    volume. Note that, since the effective volumes are not mutually exclusive, the total curve is not the sum of all the others.
    Decay triggers are negligible for muons, so the effective volumes containing decays and the 'No FI' volume are ignored.
    Bottom panel: ratio of the total effective volume over the first interaction effective volume.}
    \label{fig:muon_volumes_1}
\end{figure}

\begin{figure}
    \centering
    \includegraphics[width=0.48\textwidth]{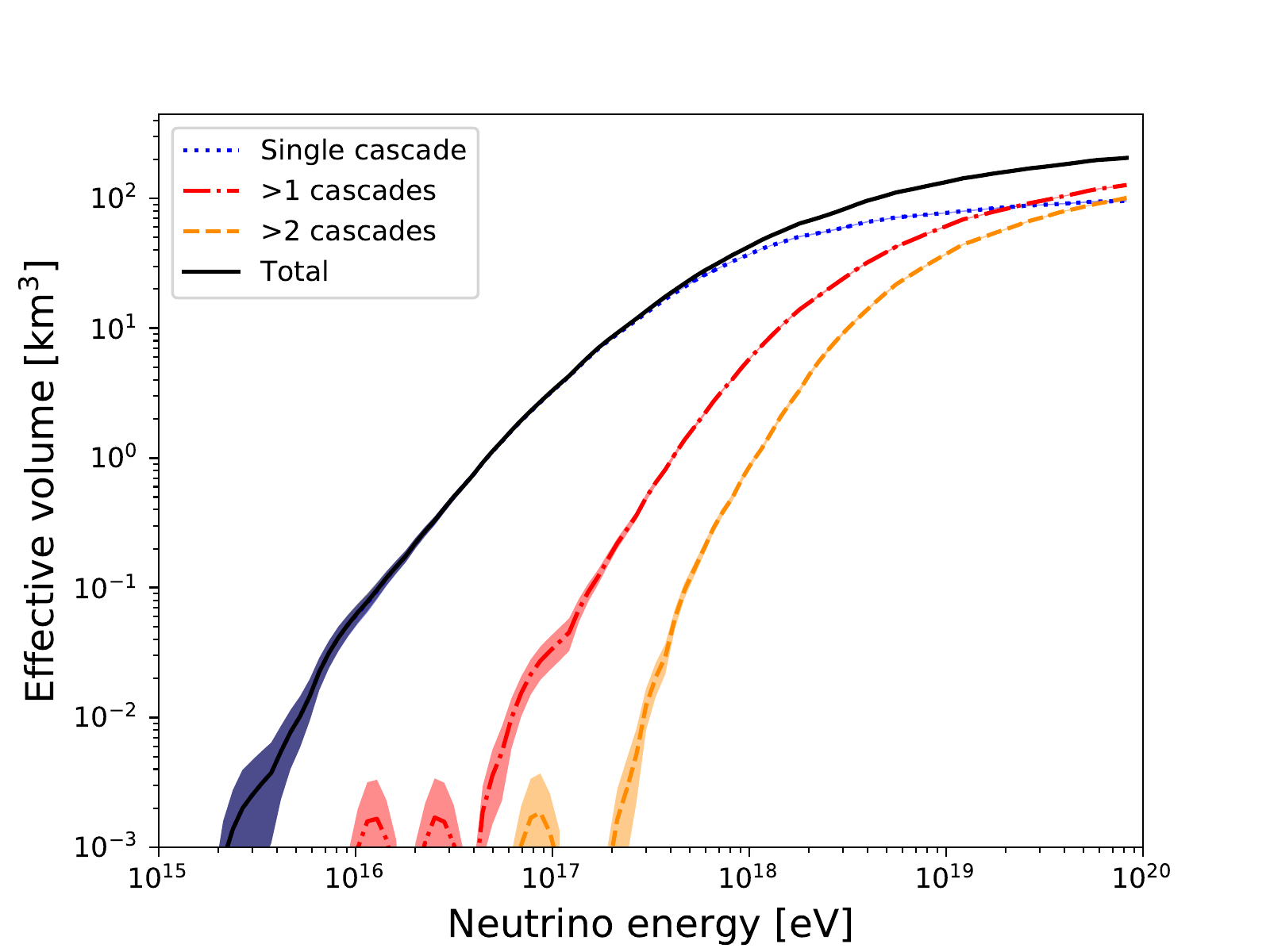}
    \caption{Same as Fig.~\ref{fig:muon_volumes_1}, but with different types of effective volumes: single, events triggered by one interaction (cascade) only; multiple cascades, event triggered by more than one interaction; and $>2$ cascades,
    event triggered by more than two interactions. The total effective volume is also shown.}
    \label{fig:muon_volumes_2}
\end{figure}

In Fig.~\ref{fig:muon_volumes_1} and \ref{fig:muon_volumes_2} we can find the effective volumes for muon neutrinos. As muons generally have low energies upon decay, the
decay-induced showers are not detectable with radio. The effective
volume for muon radiative losses is larger than the tau equivalent, which is expected as muons radiate more than
taus. The chance of detecting the first neutrino interaction and a muon radiative loss from the same event is larger for muons at low
energies. However, this does not imply that a double cascade at low energies comes from a muon because the taus compensate this
deficit with the effective volume of decay, hindering in principle the possibility of muon/tau distinction with low-energy double cascades.
However, there is some room for exploring it at high energies.
We show in the bottom panel of Fig.~\ref{fig:muon_volumes_1} the relative increase in muon effective volume with respect to the
first interaction only case. Below \SI{100}{PeV}, the increase is $\sim 40$\%, smaller than for the tau neutrinos. However, around
EeV, it reaches $\sim 50$\%, more than in the tau neutrino case. The ratio then decreases slowly with increasing energy, arriving
at $\sim 25$\% at \SI{100}{EeV}.

\begin{figure}
    \centering
    \includegraphics[width=0.48\textwidth]{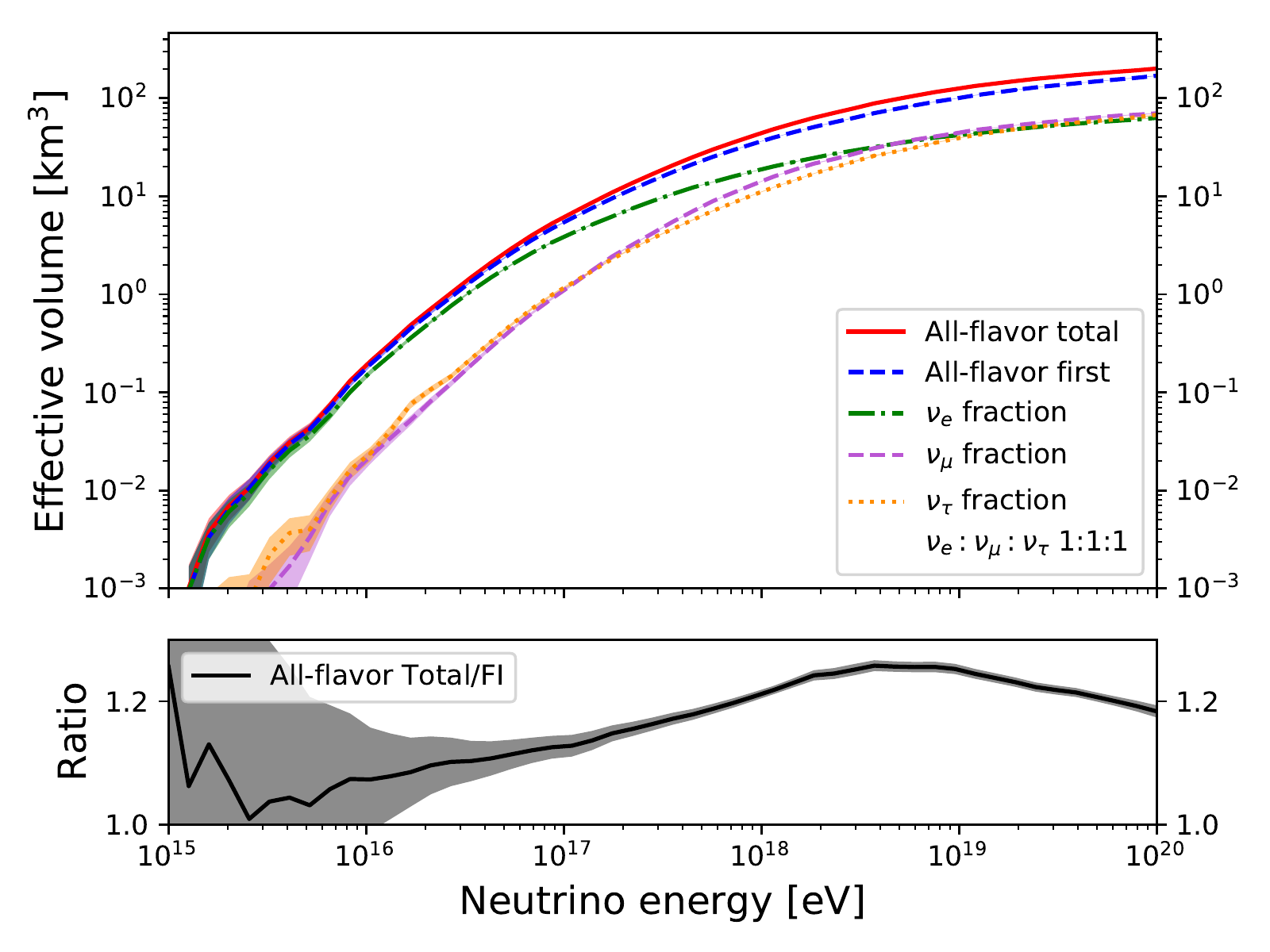}
    \caption{Top panel: all-flavor neutrino effective volumes for an $10\times 10$ square array of \SI{100}{m} deep dipoles, with $1.5\sigma$ threshold.
    The shades represent the $1\sigma$ uncertainty assuming a poissonian distribution. Total and first interaction
    (FI) volumes are shown. The fractional contributions for each flavor (assuming a 1:1:1 flux) are also shown.
    Bottom panel: ratio of the total effective volume over the first interaction effective volume.}
    \label{fig:total_ratio}
\end{figure}

For designing experiments, it is interesting to investigate how much these secondary interactions increase the all-flavor neutrino effective volume, and as a
consequence, the chances of detecting any neutrino. To this end, also the electron neutrino effective volume for the $10\times 10$ array has
been calculated, which is not affected by secondaries, however contributes a comparatively larger fraction of the effective volume, as typically all products of NC and CC interactions are detectable. Assuming a 1:1:1 ratio for the incoming flux, we show the total
and first interaction effective volume in Fig.~\ref{fig:total_ratio}. The increase in effective
volume grows with neutrino energy up to \SI{10}{EeV}, reaching $\sim 25$\%, and then decreases with energy. The total increase
ranges between 20\% and 25\% above \SI{10}{PeV}. This increase is due to the secondary interactions
from muon and tau neutrinos only, since electron neutrinos are not subject to secondary interactions. 

\subsection{Number of detected interactions per particle}

\begin{figure}
    \centering
    \includegraphics[width=0.48\textwidth]{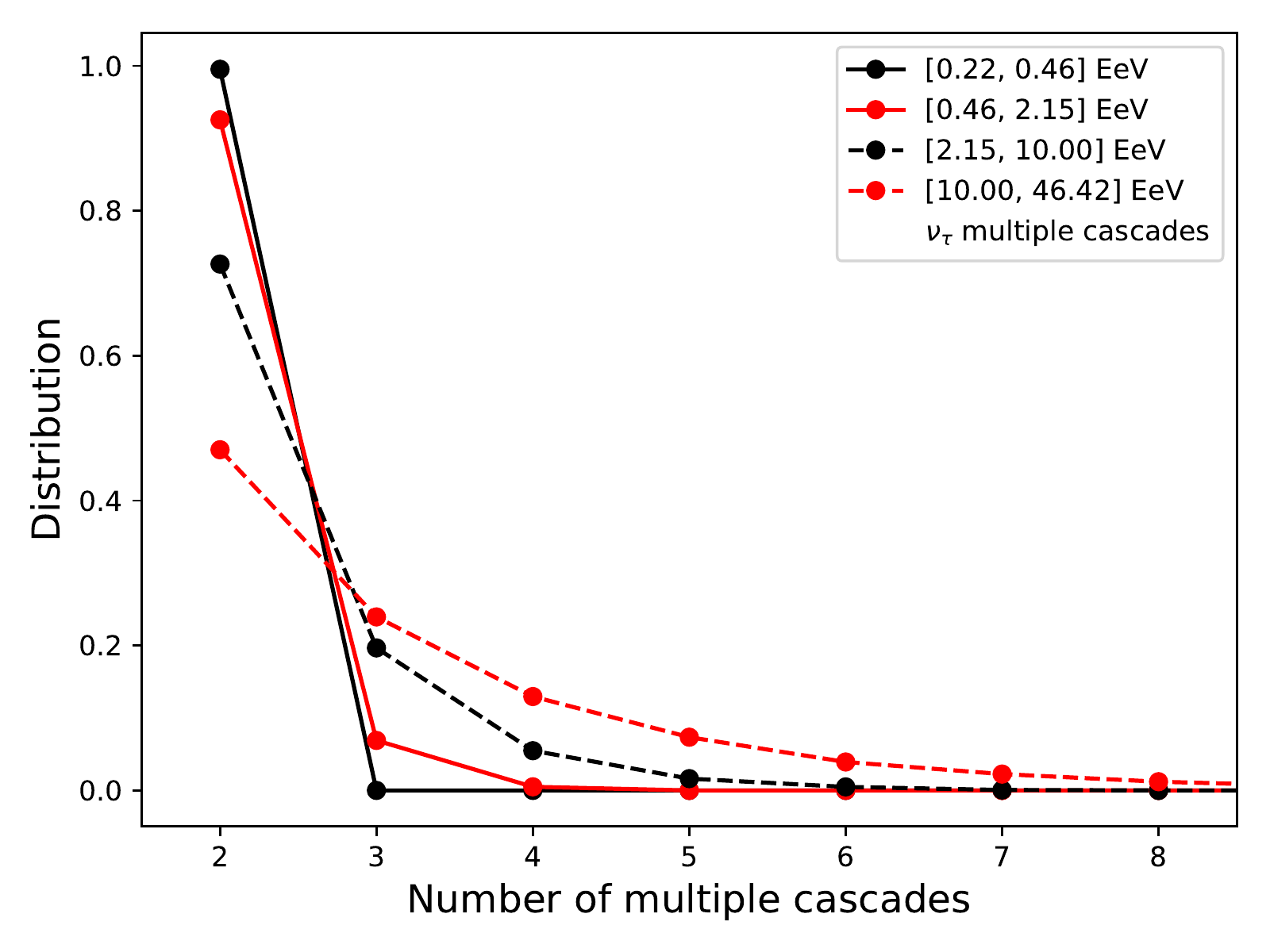}
    \includegraphics[width=0.48\textwidth]{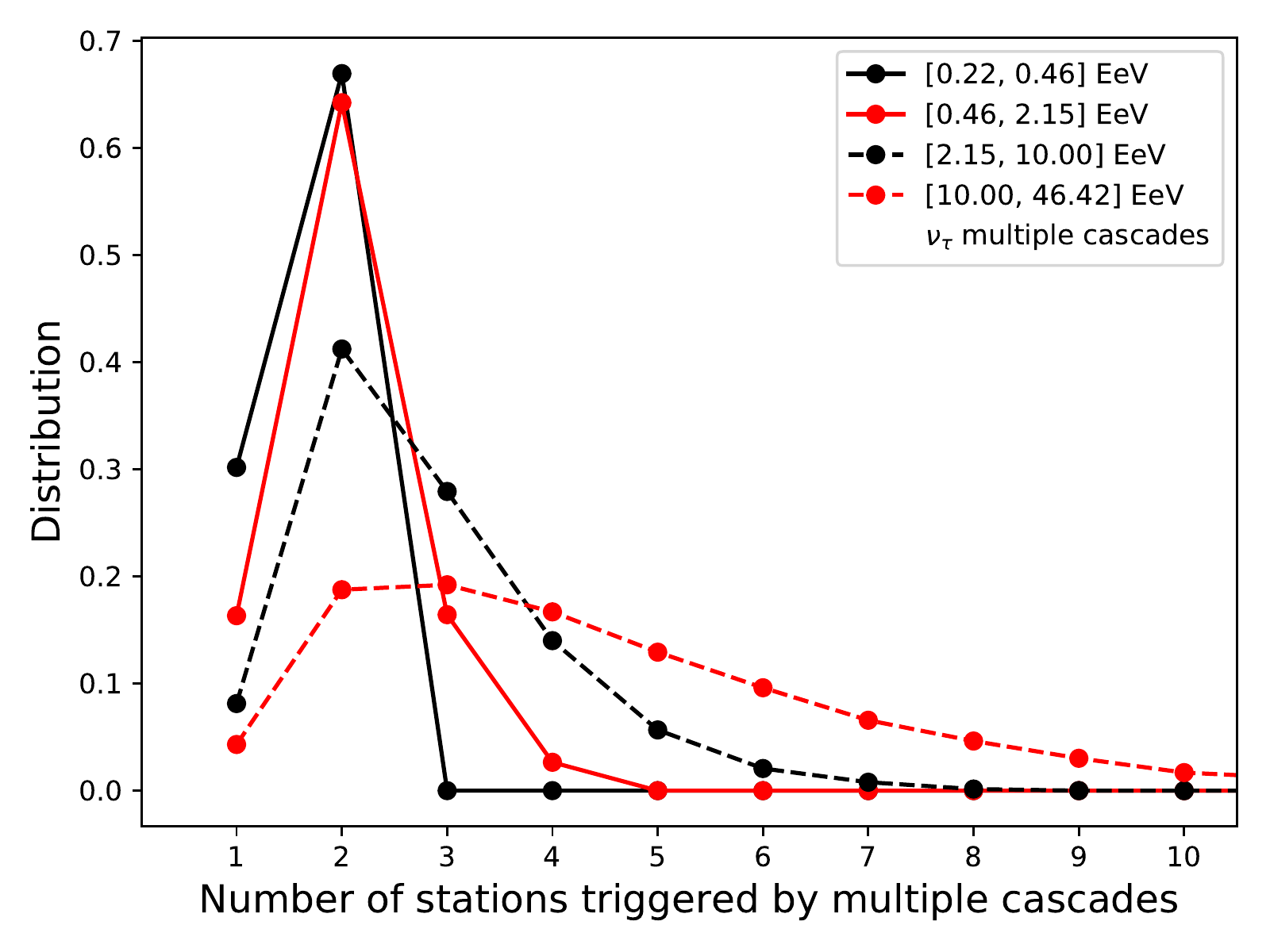}
    \caption{Top: distribution of the number of multiple cascades induced by tau neutrinos and detected by a $10\times 10$ dipole array for several
    neutrino energy bins. Bottom: same as top, but with the number of stations triggered by multiple-cascade events.}
    \label{fig:tau_mult}
\end{figure}

We also calculate how many multiple signatures are detected by a $10\times 10$ dipole array. We show in Fig.~\ref{fig:tau_mult} (top) the
distribution of the number of multiple-cascade events created by a tau neutrino that trigger the array.
A multiple-cascade event is defined as an event containing multiple interactions from the same parent neutrino that trigger the detector more than once.
Therefore, a double-cascade event (triple, etc) is an event presenting two (three, etc) showers that trigger a detector.
Curves for several energy bins are depicted in Fig.~\ref{fig:tau_mult}.
For the lowest energy bin on the plot ($0.22$ to $0.46$ EeV), at most only two cascades can be detected. As the energy increases,
detection of more than two cascades becomes possible, and at tens of EeV, $\sim 55\%$ of the times an event with more than two triggering cascades is detected.
The average multiplicity is more than two, which means it is more likely to detect more than two cascades.

A related question can be asked: how many stations are triggered by multiple-cascade tau neutrino events? We show in
Fig.~\ref{fig:tau_mult} (bottom) the distribution of the number of stations triggered by multiple tau-neutrino-induced cascades. 
At low energies, the number of triggered stations by multiple-cascade events is two in most of the cases. However, if two showers are created really close, several cascades can be detected with a single station.
With increasing
energy, the expected number of stations goes up as one would expect, to the point that at tens of EeV the distribution shows
a really long tail and events with more than 5 stations are not unlikely. Such events would constitute a characteristic signature
for high-energy neutrinos.

\begin{figure}
    \centering
    \includegraphics[width=0.48\textwidth]{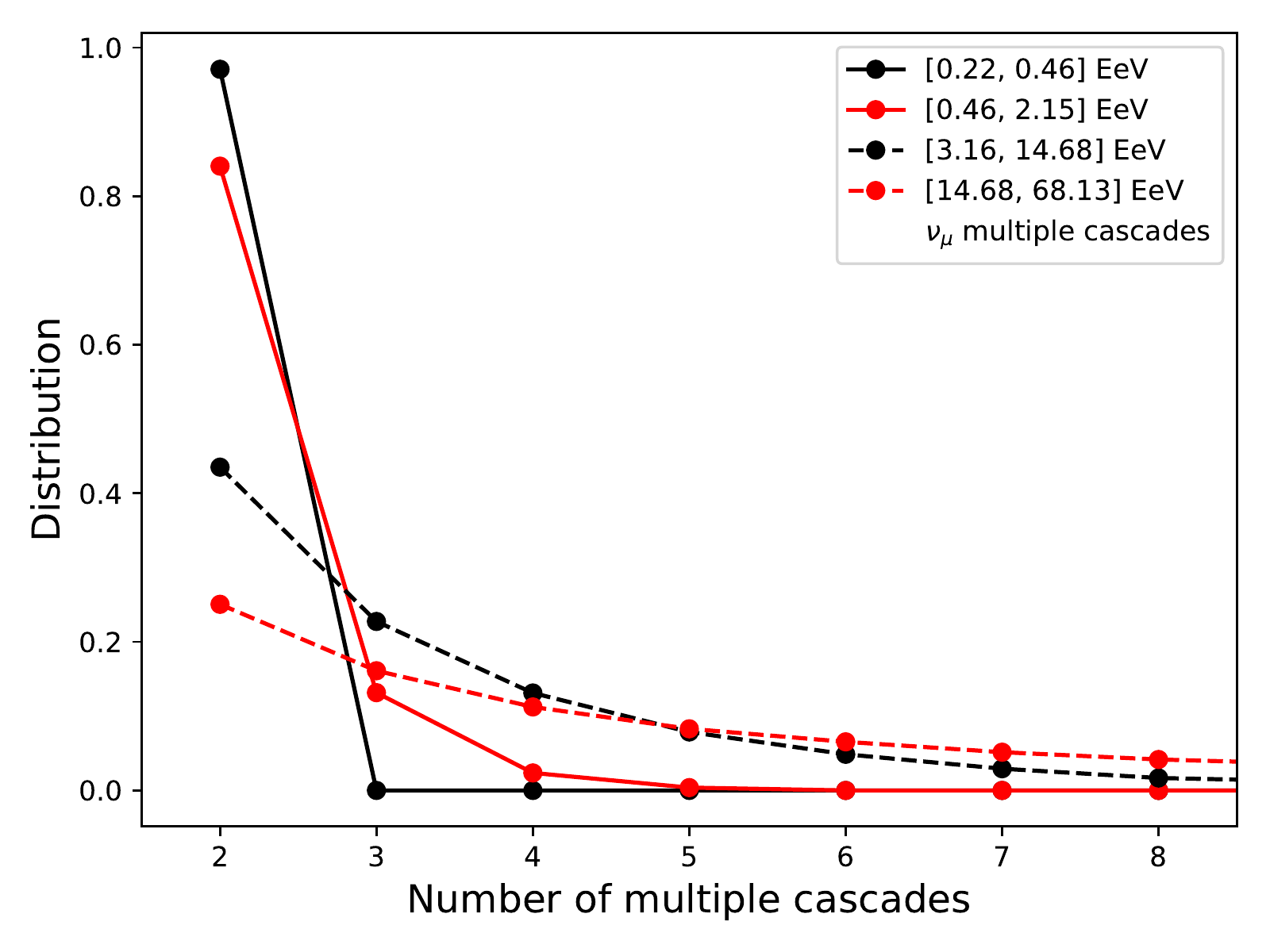}
    \includegraphics[width=0.48\textwidth]{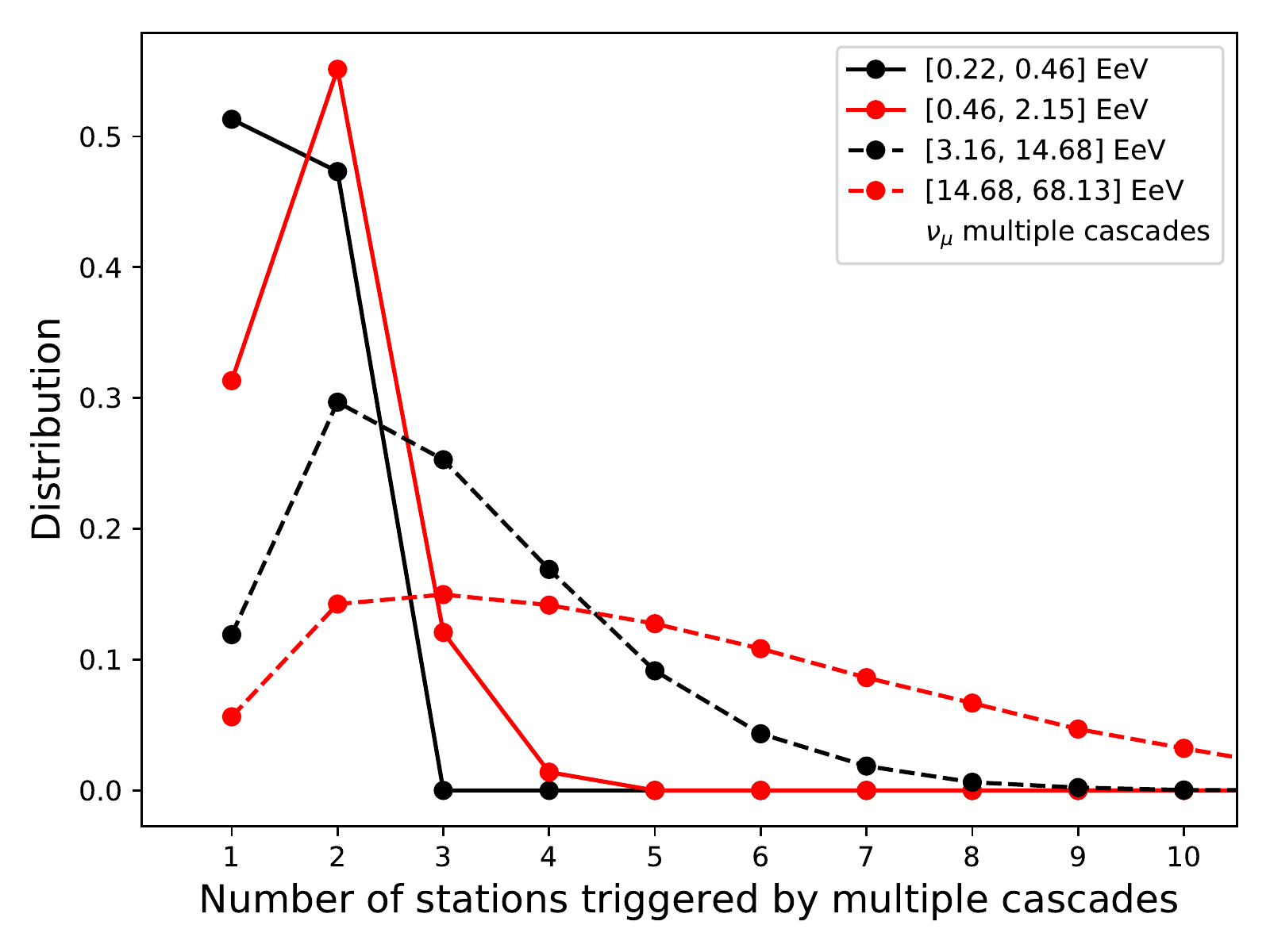}
    \caption{Same as Fig.~\ref{fig:tau_mult} but for muon neutrinos.}
    \label{fig:mu_mult}
\end{figure}

For Fig.~\ref{fig:mu_mult}, we repeated the same analysis for muon neutrinos. Muons radiate more showers than taus,
which lets them create more detectable cascades than taus and illuminate more stations. However, both particles present similar
distributions and can produce events with large multiplicity at high energies.

It is worth noting that the distributions presented in this work are valid for a detection threshold of $1.5\sigma$, so the distributions are expected to change if the detection threshold changes.

\subsection{Distance and time difference distributions for tau-neutrino-induced double-cascade events}

\begin{figure}
    \centering
    \includegraphics[width=0.48\textwidth]{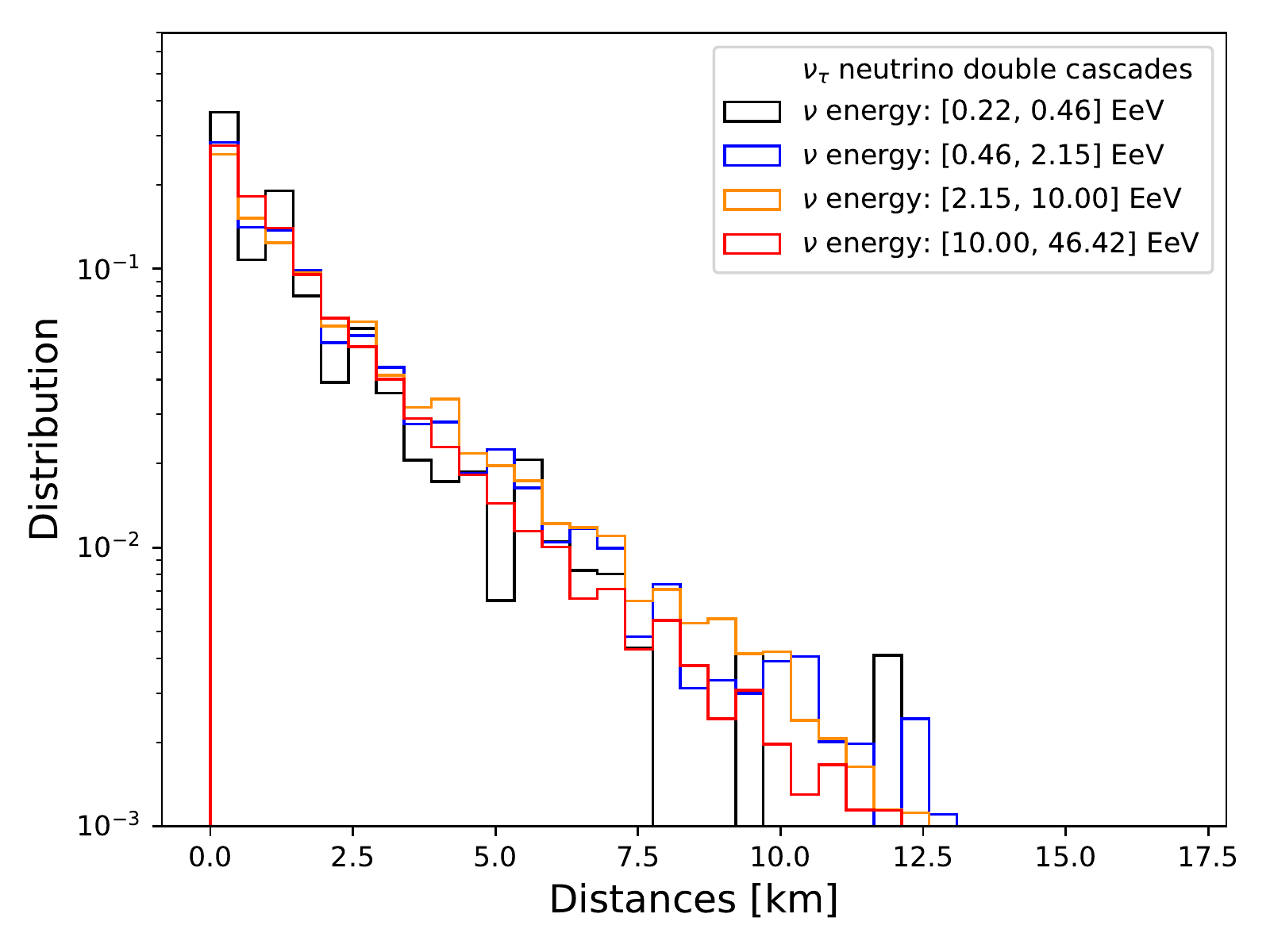}
    \includegraphics[width=0.48\textwidth]{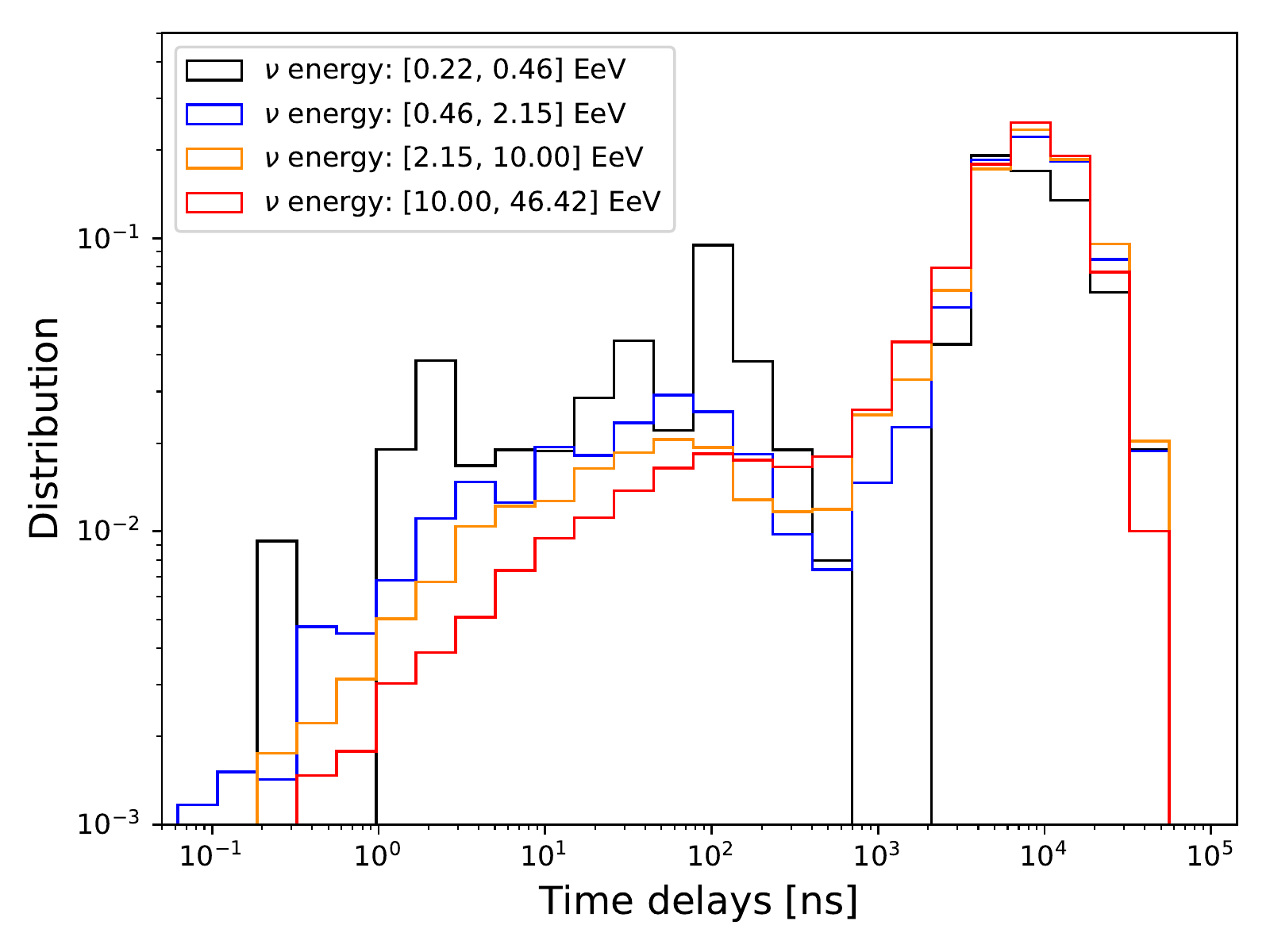}
    \caption{Top: distribution of the distances between cascades of double-cascade tau neutrino events detected by a $10\times 10$ dipole array for several
    neutrino energy bins. All the distributions are normalized to 1. Bottom: same as top, but with the signal arrival times at the station. The antenna distance between nearest neighbors is \SI{1.25}{km} for this study.}
    \label{fig:tau_dist}
\end{figure}

For optimizing the design of a detector with regard to an improved sensitivity towards multiple signatures, both the timing and the distance of the interactions are relevant. For the sake of simplicity, we will restrict this discussion to double-cascade tau neutrino events, defined as events that have
triggered the detector twice. An interesting aspect is the distribution of distances between the two interactions that
cause the triggers in the $10\times 10$ array.
The distance distribution is found in Fig.~\ref{fig:tau_dist}, top. Low energies present slightly higher probability of creating two detectable
double cascades within short distances, what agrees with the fact that at low energies a single station is likely to detect
multiple cascades. High energies also can tap into larger distances, although the overall energy dependence of the distribution is not that prominent.

\begin{figure}
    \centering
    \includegraphics[width=0.48\textwidth]{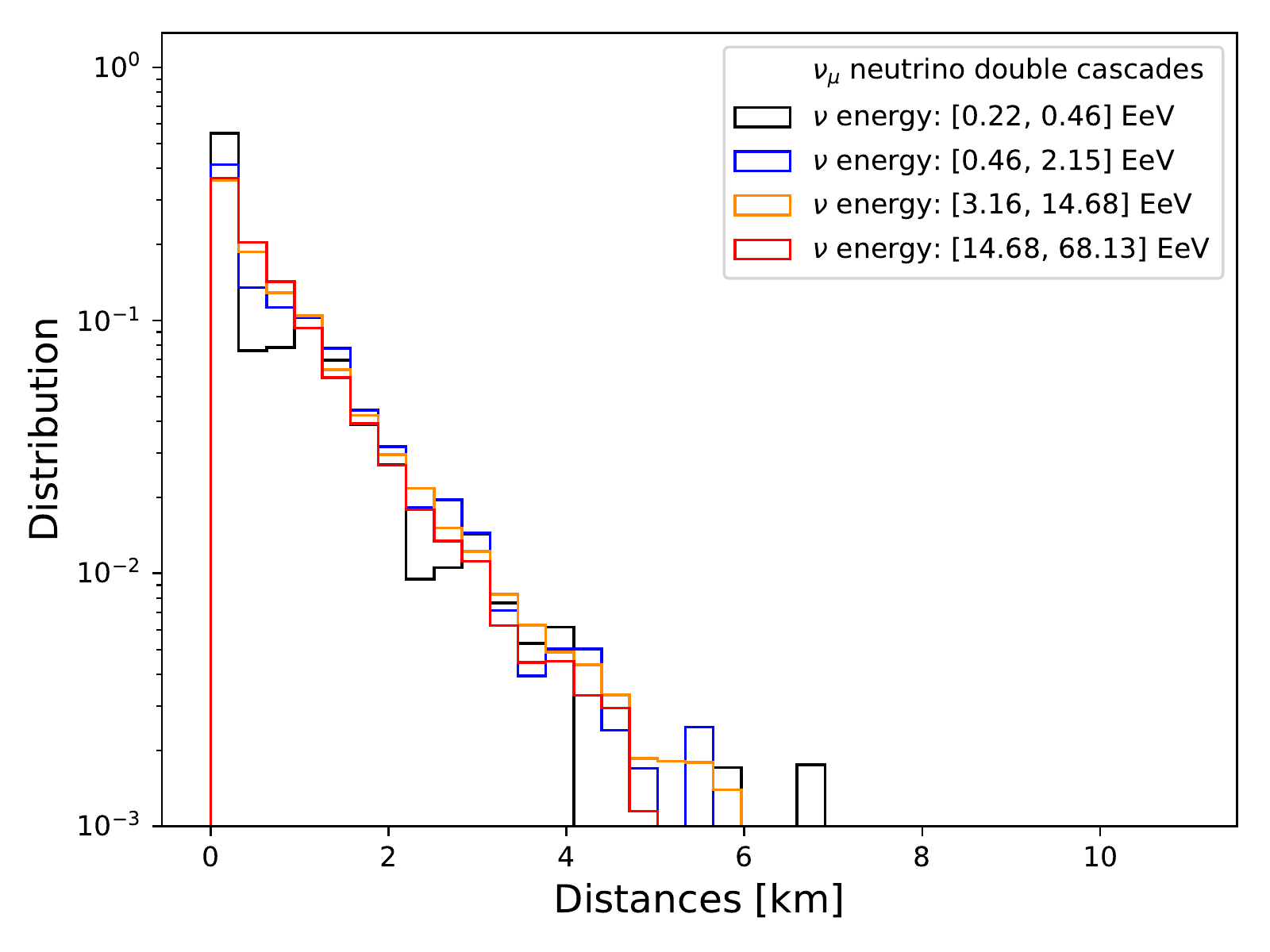}
    \includegraphics[width=0.48\textwidth]{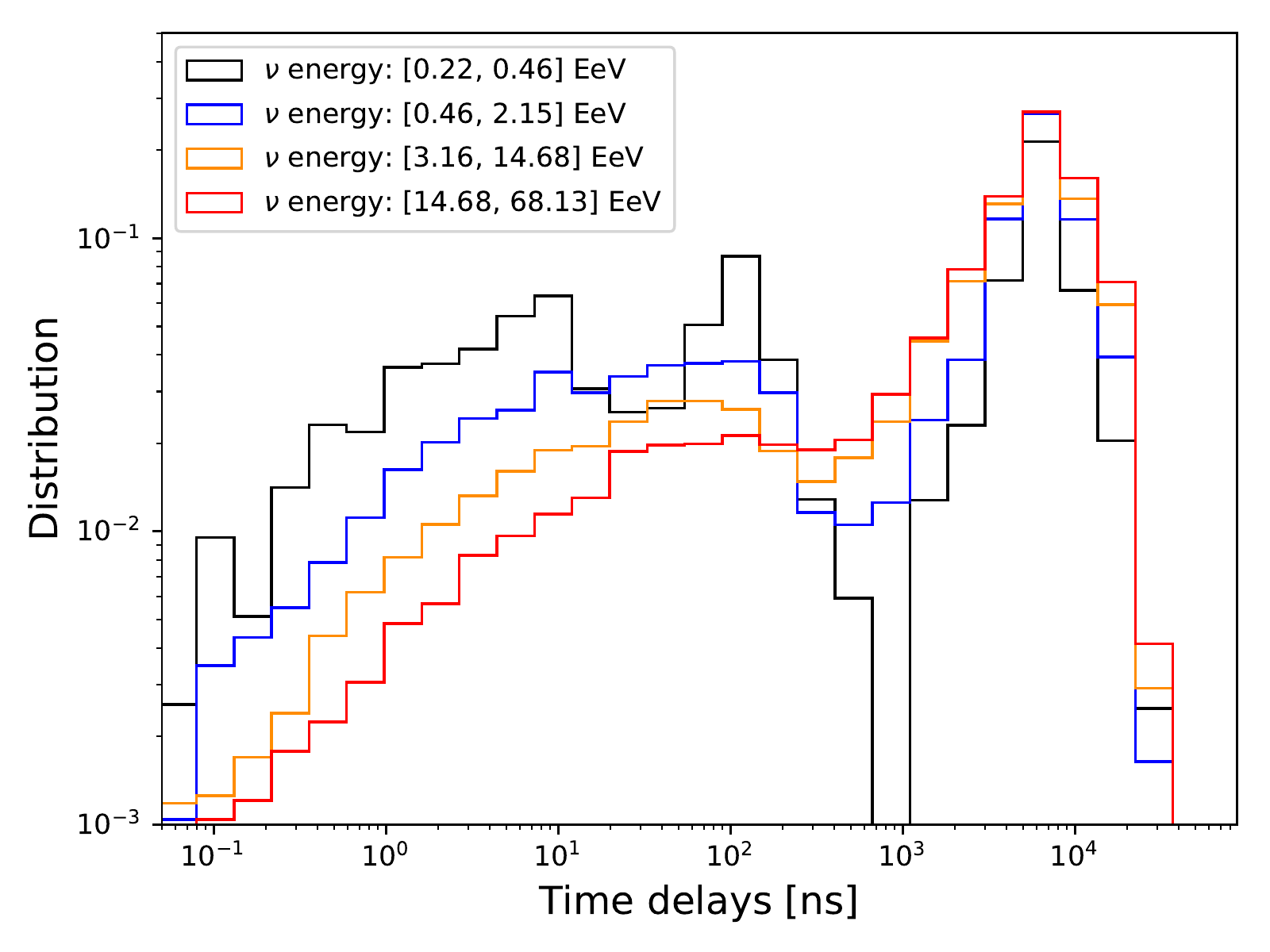}
    \caption{Top: distribution of the distances between cascades of double-cascade muon neutrino events detected by a $10\times 10$ dipole array for several
    neutrino energy bins. Bottom: same as top, but with the signal arrival times at the station.}
    \label{fig:mu_dist}
\end{figure}

We show in Fig.~\ref{fig:tau_dist}, bottom, the distribution of difference in arrival times, defined as the times when the electric field signals from each cascade arrive at the detecting stations. For each cascade, we take the ray solution (direct, refracted, or reflected) that presents the largest amplitude. The same is shown for muon neutrinos in Fig.~\ref{fig:mu_dist}.
As it was suggested by the distance distribution, low-energy taus create double cascades that lie closer in distance, while the time delay distribution suggests they also present smaller detection time differences than high-energy taus. Arrival time differences tend to be smaller for muon than for taus. Besides, distances between double cascades tend to be smaller for muons than for taus, which
is explainable because muons radiate more often than taus. We can see that there is a correlation between distances and arrival
times, as both the average distance and arrival time difference are smaller for muons.

\begin{figure}
    \centering
    \includegraphics[width=0.48\textwidth]{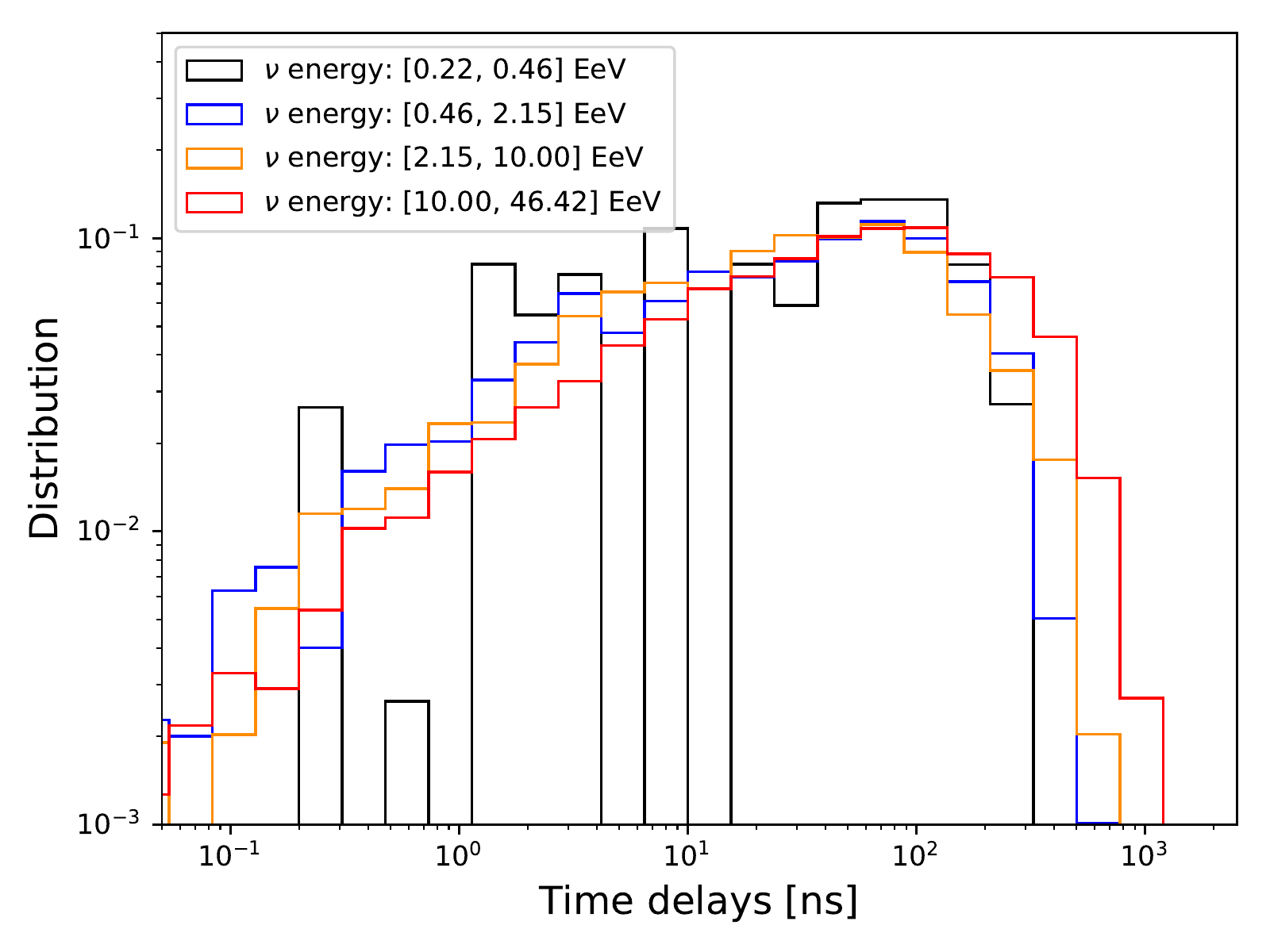}
    \caption{Same as Fig.~\ref{fig:tau_dist}, bottom, but for double-cascade events stemming from tau neutrinos detected by a single antenna. The distribution for muon neutrinos is very similar.}
    \label{fig:tau_single}
\end{figure}

Given that it is not unlikely to detect a double-cascade using a single antenna, it would be advisable to know the
arrival times distribution for a single antenna and deduce the time trace length a detected event should have so as not to miss
one of the cascades. One can see in Fig.~\ref{fig:tau_single} that the time difference distribution reaches a value of \SI{1}{microsecond}
in a limited number of cases, so a trace of a few microseconds could, in principle, capture these single-station double-cascade
events. The distribution implies, however, that a non-negligible number of shower signals arrive within
few tens of nanoseconds, which means they will most likely interfere, depending on the bandwidth and group delay
characteristics of the detector. This interference from different showers is currently not taken into account but should be incorporated in future simulations, since it can
be either constructive or destructive and might modify slightly the effective volumes and trigger distributions.

Using a typical radio band for in-ice neutrino detection (for instance, from \SI{\sim 100}{MHz} to \SI{\sim 700}{MHz}, as it is 
expected for RNO-G) and including the group delay induced by the antenna and electronics, we expect the detected neutrino pulses
from a single shower (without LPM) to be shorter than \SI{\sim 10}{ns}, which means that interference between pulses will happen
if and only if the difference in arrival time of the pulses is less than \SI{\sim 10}{ns}.
Judging by Fig.~\ref{fig:tau_single}, $\sim 30\%$ of all tau double-cascade pulses are detected by a single antenna with a difference of less than \SI{10}{ns}. This ratio has a weak dependence on energy. However, we have to factor in that the number of double-cascade pulses seen by a single antenna is less than the total number of double-cascade pulses seen. After correction, we obtain that at hundreds of PeV $\sim 12\%$ of tau double-cascade events present interference, and as energy increases this number goes down, reaching $\sim 3\%$ at tens of EeV. The numbers are a slightly higher for muons: $\sim 30\%$ at hundreds of PeV, and $\sim 5\%$ at tens of EeV
As a result, we can expect interference to be significant for $\sim 4\%$ percent of the double-cascade events at tens of EeV, while at hundreds of PeV the ratio could be of $\sim 15\%$, assuming a 1:1:1 flavor ratio.

\subsection{Neutrino flavor sensitivity}

It would be interesting to distinguish flavor using these and other signatures, especially to study fundamental physics with radio neutrino detectors \cite{Ackermann:2019cxh}. Electron neutrinos, for instance, create a
hadronic shower and an electromagnetic shower at roughly the same location after undergoing a CC interaction. These two showers are
expected to be detected as a single cascade. Only at high energies (above \SI{\sim 1}{EeV}) the LPM effect is relevant and the 
electromagnetic shower is delayed and can show multiple spatially displaced sub showers (see e.g.\ examples in \cite{NuRadioMC}). These cascades are separated tens of meters at most and if detected would provide a signature for electron neutrinos. This also means that if one were to detect two (or more)
cascades at distances of hundreds of meters or kilometers, the initial flavor was either
muon or tau neutrino. 

In Fig.~\ref{fig:muon_volumes_2}, we can see that the muons at high energies can produce multiple cascades at a rate comparable
to single cascades. The probability of detecting more than two cascades is larger than for the tau case and in fact it reaches the
effective volume for a single cascade at \SI{\sim 5e19}{eV}. A muon is more likely to create multiple-cascade events than a tau, and is also more
likely to produce more cascades on average than a tau.

This can be used in conjunction with the different distances and wave arrival times between cascades (see 
Figs.~\ref{fig:tau_dist} and~\ref{fig:mu_dist}).
The idea is to create a set of observables such as number of cascades, distances between cascades, arrival times, electric field amplitudes,
reconstructed shower energy, etc.\ and calculate the probability of measuring this set of observables assuming a fixed input flavor.
Then, a Bayesian analysis could be used to estimate the probability that an event with some measured values for this set has been
initiated by a specific neutrino flavor, similarly to what the IceCube collaboration does for flavor sensitivity \cite{Stachurska:2019wfb}. 
A rigorous Monte Carlo study is needed to know the feasibility of such an approach, but these results indicate a strong possibility. 

\section{Conclusions}

We have shown in this article that a proper treatment of lepton propagation in dense media such as ice is crucial for
the correct simulation of the sensitivity and performance of radio neutrino experiments. High-energy taus and muons create
showers during their propagation and if the energy is large enough they are detectable by a radio array. Tau-induced showers, when the tau energy is below 
\SI{\sim 20}{PeV}, stem mainly from decays and can be either hadronic or electromagnetic. When the tau energy is higher, pair production dominates the shower creation and produces electromagnetic showers. However, photonuclear interaction, which is the subdominant process, creates hadronic cascades with a much higher average energy than the pair-production-induced showers. Muons radiate mainly via bremsstrahlung below \SI{\sim 100}{PeV}, which creates electromagnetic cascades,
and via pair production and photonuclear interaction at higher energies, in a similar way to taus.

We have studied the influence of the atmospheric muon background on an in-ice radio array. Assuming a downward-going atmospheric
muon flux, we have propagated these muons in ice with the lepton propagation code PROPOSAL and quantified the response of a radio array to these muons in the
form of an effective area. Using MCEq, the muon flux at the surface was calculated and combined with this effective area to
arrive at the number of muons detected by a 100-station radio array. The expected numbers are, for a $1.5\sigma$ dipole array,
between 0.43 and 2.19 muons per year, where the large uncertainty comes from the hadronic interaction models. These numbers do not include the uncertainties of the assumed cosmic ray flux model. The precise numbers are also very sensitive to the trigger that is used. For example, for a trigger like the phased array trigger envisioned for the RNO-G experiment, we expect an order of magnitude less, between 0.048 and 0.311 muons to trigger per year. Muon-induced background also decreases when the antennas are closer
to the surface, and in principle many events could be vetoed by a surface array sensitive to the radio emission of the air shower. This is because most background muons that are observed by an in-ice detector are relatively inclined
and the associated air showers have energies above a few tens of PeV, which is coincidentally the threshold for a sparse air-shower radio array. Atmospheric muon background is not easily distinguished from neutrino signals looking at vertex position and arrival direction alone, but several variables may be used to distinguish it on a statistical basis. Measuring the shower energy may allow improving of the signal-to-background ratio which increases with shower energy for most neutrino-flux expectations. 

We have furthermore calculated the effective volumes of a square $10\times 10$ array for muon and tau neutrinos. If the initial neutrino interacts
via charged current, the outgoing lepton can create additional showers that can be detected, sometimes together with the first interaction, sometimes
isolated with no first interaction counterpart. These additional interactions add to the effective volume of a neutrino detector. For tau neutrinos the correction is strongest at the PeV scale with an additional 60\%, while at high energies the correction is roughly 25\%. For muon neutrinos, the correction is largest around one EeV, where it reaches 50\%, and then at both sides of the spectrum goes down to 25\% around one PeV and hundreds of EeV.
Assuming a 1:1:1 flavor ratio, these effects add between 10 and 25\% to the total neutrino effective volume.

We have also investigated the number of stations that are illuminated by multiple showers coming from the same parent neutrinos. Due to the
random nature of the stochastic losses, if two of them occur close or if one of them occurs closer to the first neutrino interaction,
both interactions can be seen by a single detector station, and in fact this is a likely channel up to EeV energies. Above EeV energies, this probability
decreases. From hundreds of PeV to tens of EeV, the most likely number of detector stations lit by multiple interactions is two, but at higher
energies the mode of the distribution becomes larger and its shape becomes increasingly flatter, and the detection of an event in
eight or more antenna stations in a $10\times 10$ square array becomes as likely as detection by a single antenna. Muons and taus present
similar distributions.

We have also studied in particular the distribution of distances between interaction vertices and their timing for detected double cascades (two showers detected by the array). Muon double cascades tend to happen closer than tau double cascades, since muons radiate more often than
taus. The timing difference imposes some constraints on hardware design if one wants to capture both pulses with a single
trigger. We have found that it is rare for these events to arrive with a time delay of more than \SI{1}{microsecond} with respect to each other up to tens of EeV. So experiments either have to store sufficiently long records or allow for double buffering of signals within this time-frame. 

A correct treatment of lepton propagation is essential for the analysis and reconstruction of the neutrino-induced signals that the next generation of radio-based
neutrino detectors will attempt to measure. As it is the standard for optical neutrino telescopes, a flavor sensitivity of radio neutrino experiments is highly desirable and should be included in design decisions.  

\section*{Acknowledgements}
The authors would like to thank their colleagues from the radio neutrino community for the lively discussions and input on this manuscript. We would also like to thank the authors of PROPOSAL, especially Jan Soedingrekso, for cross-checking the validity of the code at higher energies and helping us getting started. We thank Anatoli Fedynitch for useful discussions about hadronic and cosmic-ray flux models. The authors acknowledge funding from the German Research Foundation (DFG) under grant NE 2031/2-1 (DGF, AN) and GL 914/1-1 (CG).

\appendix

\section{Atmospheric muons with realistic triggers} 
\label{sec:triggers}

The location of the experiment, the detector layout and the chosen trigger strongly influence the detected muon background. A systematic comparison of all factors is beyond the scope of this article. However, for reference purposes we would like to discuss two trigger configurations close to what has been proposed as experimental set-ups. We hope that this will illustrate the complexity involved. 

The first scenario consists of a 4-antenna phased array at
\SI{\sim 100}{m} depth with 30 phasing directions from $50$ to $-50$ degrees of elevation. The signals for each channel are
filtered using a diode and the amplitude threshold is set such that the noise trigger rate is \SI{\sim 1}{Hz} at a noise temperature of \SI{300}{K} in the band of \SIrange{132}{700}{MHz}. This configuration
is set in a medium modeled after Summit Station, Greenland, with a \SI{3}{km} deep layer of ice. This configuration
has been inspired by the \hbox{RNO-G} project \cite{Aguilar:2019jay}, which will start installation in Greenland in summer 2021.

The second one consists of 4 downward-pointing log-periodic dipole antennas (LPDA) triggered with a 2 out of 4 coincidence scheme. The noise trigger
rate has been chosen to be \SI{\sim 10}{mHz} at a noise temperature of \SI{250}{K} in the band of \SIrange{80}{150}{MHz}. These antennas are located \SI{3}{m} beneath the surface of the Ross Ice Shelf in Antarctica,
where the ice layer is only \SI{550}{m} thick, exposing less volume for incoming particles to interact with, but where the
electric field can be reflected on the bottom layer and detected by the surface antennas \cite{glaser_icrc}. This configuration
has been inspired by the ARIANNA experiment on the Ross Ice Shelf \cite{Anker:2019mnx}. Currently, there is a proposal under discussion to expand this site to 200 stations \cite{2020ARIANNA-200}. 

\begin{figure}
    \centering
    \includegraphics[width=0.48\textwidth]{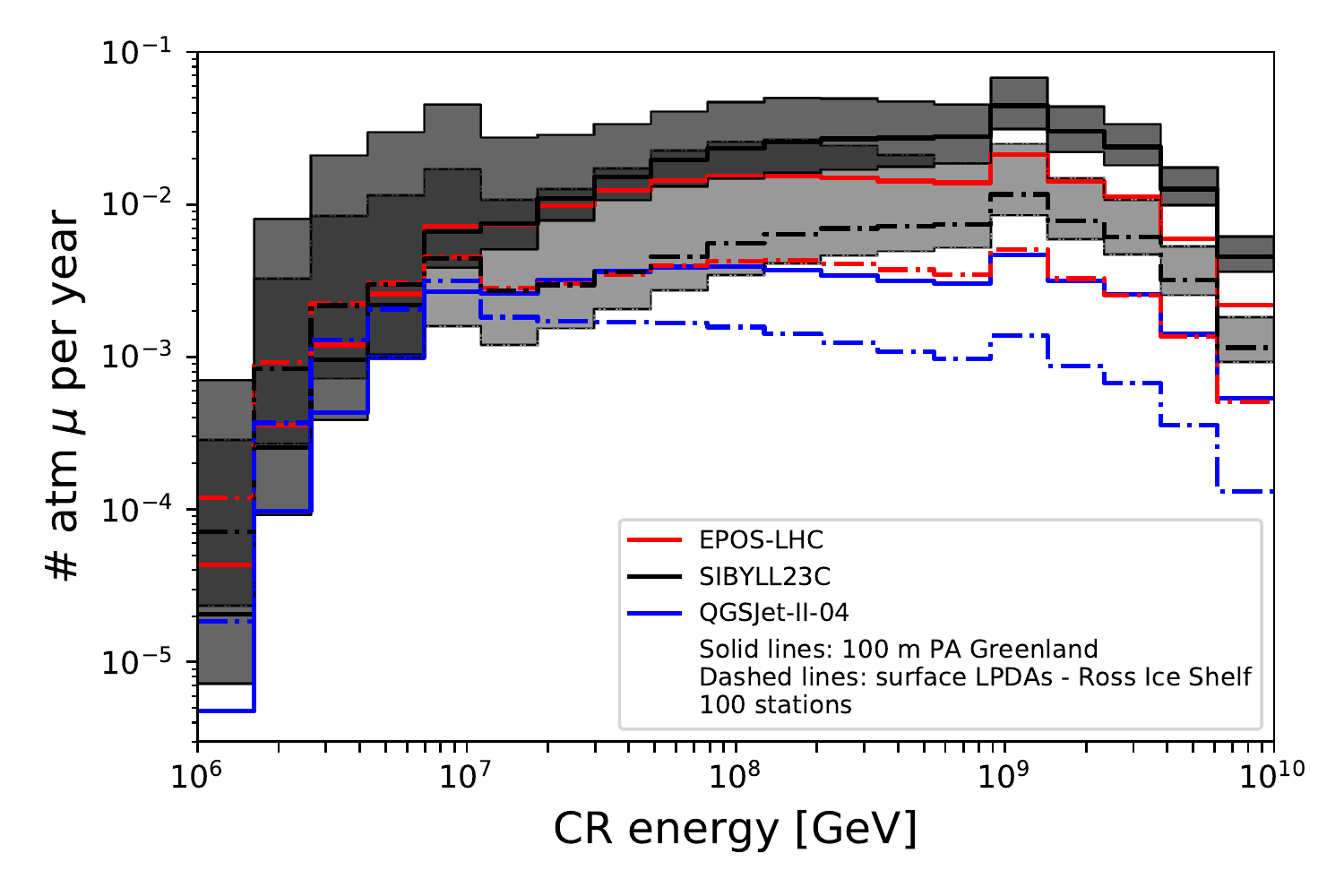}
    \includegraphics[width=0.48\textwidth]{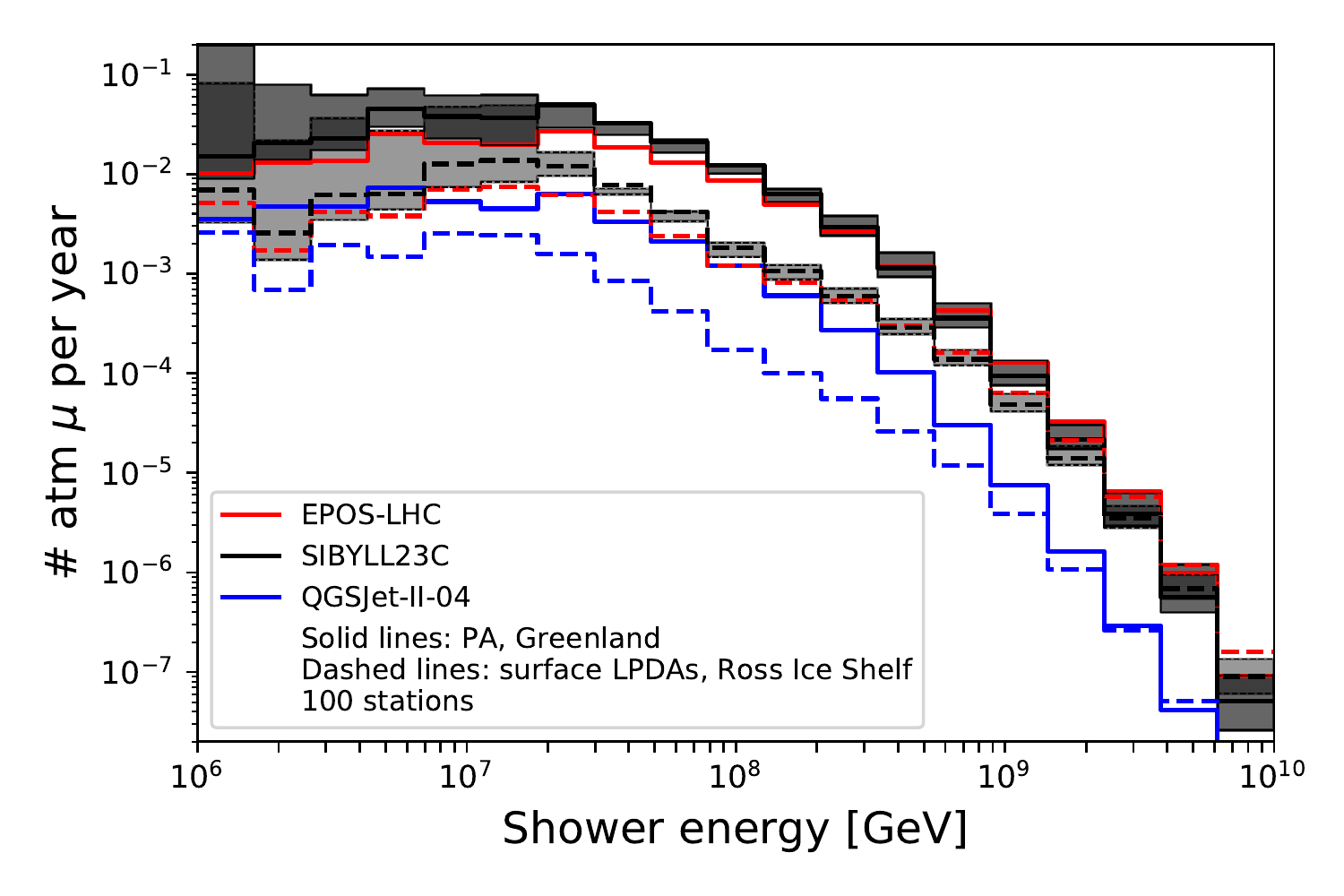}
    \caption{Top: histograms containing the average number of atmospheric muons detected by a 100-station array as a function of
    cosmic ray energy in GeV for several hadronic models. Numbers for 100 independent envelope dipole phased arrays are shown (\SI{\sim 1}{Hz} noise trigger rate), located near Summit Station, Greenland, as well as surface LPDA trigger (\SI{\sim 10}{mHz} noise trigger rate) on the Ross Ice Shelf. 
    The shaded bands represent the uncertainties induced by the cosmic-ray flux model, the hadronic model, and the effective area calculation, for the SIBYLL 2.3C model only.
    See text for details. Bottom: same as top, but as a function of muon shower energy.}
    \label{fig:triggers}
\end{figure}

The results for the yearly numbers of detected atmospheric muons can be found in Fig~\ref{fig:triggers}.  The corresponding yearly numbers can be found in Table~\ref{tab:muon_numbers_realistic}.

\begin{table}[b]
\begin{center}
\begin{tabular}{ |c|c|c| }
 \hline
                 & PA \SI{100}{m} & LPDAs \\ \hline \hline
 SIBYLL 2.3C     & \SI{0.311}{}  & \SI{0.088}{}  \\ \hline
 EPOS-LHC        & \SI{0.185}{}  & \SI{0.057}{}  \\ \hline
 QGSJet-II-04    & \SI{0.048}{}  & \SI{0.010}{} \\ \hline
\end{tabular}
\caption{Number of detected atmospheric muons per year 
for a 100-station array. Three hadronic models and two detector layouts are shown. PA stands for the phased array at
\SI{100}{m} of depth near Summit Station, while LPDAs stands for the surface LPDAs antennas on the Ross Ice Shelf. 
The relative uncertainties due to cosmic ray flux, hadronic modeling, and effective area are similar across models. The uncertainty on the detected muon numbers for the SIBYLL 2.3C model are $\sim {}^{+0.4}_{-0.1}$ for the phased array, and $\sim {}^{+0.20}_{-0.04}$ for the surface LPDAs.
See text for
details.}
\label{tab:muon_numbers_realistic}
\end{center}
\end{table}

The number of
triggered background events is significantly lower for the LPDAs on the Ross Ice Shelf.
However, the reduced number of background events alone should not be used to compare the suitability of the designs. It is foreseen that both experimental set-ups will have a cosmic-ray self-veto, which will reduce the background by tagging the air showers from which the muons stem. The efficiency of such a veto will also depend in detail on station spacing, height above sea-level, antenna type and orientation, geomagnetic field, system noise level, and trigger algorithm. So the number of real background events, may yet be different.

Apart from logistical, financial and other practical considerations, which should be discussed elsewhere, one also needs to consider the background with respect to the expected numbers of neutrinos as in the end the number of signal events over background will be relevant. 
Therefore, we calculated the signal-to-background ratio for a variety of neutrino flux models and found that both designs have a similar ratio around \SI{1}{EeV}. The shallow Ross Ice-Shelf design performs better at lower energies and the deep Greenland design better at higher energies. 

We stress that given the flux uncertainties and the strong dependence on the experimental details the numbers are not dependable enough to prefer one approach over the other. Further studies are needed that include at least the effect of a cosmic-ray veto and possibly event reconstruction to asses the severity of the background for neutrino detection in a real detector.

\bibliographystyle{apsrev4-2}
\bibliography{bib}

\end{document}